\documentclass{pas}

\usepackage{multirow}

\usepackage[T1]{fontenc}
\begin{document}

\lefttitle{Publications of the Astronomical Society of Australia}
\righttitle{Cambridge Author}

\jnlPage{1}{4}
\jnlDoiYr{2021}
\doival{10.1017/pasa.xxxx.xx}

\articletitt{Review}

\title{Progress in quantum telescopes}

\author{David R. Gozzard$^{1,2}$, Fiona H. Panther$^{2,3}$ and Andrew G. White$^{4}$}

\affil{$^1$International Centre for Radio Astronomy Research, The University of Western Australia, Crawley WA 6009, Australia,\\
$^2$School of Physics, Mathematics and Computing, The University of Western Australia, Crawley WA 6009, Australia,\\
$^3$OzGrav: The Australian Research Council Centre of Excellence for Gravitational-wave Discovery, WA 6009, Australia and\\
$^4$School of Mathematics and Physics, University of Queensland, QLD 4072, Australia}

\corresp{D. R. Gozzard, Email: david.gozzard@uwa.edu.au}

\citeauth{Gozzard D. R., Panther F. H. and White A. G., Progress in quantum telescopes. {\it Publications of the Astronomical Society of Australia} {\bf 00}, 1--12. https://doi.org/10.1017/pasa.xxxx.xx}

\history{(Received xx xx xxxx; revised xx xx xxxx; accepted xx xx xxxx)}

\begin{abstract}
Classically, the angular resolution of a telescope is limited by its diffraction limit: the ratio of its observing wavelength to its diameter. Developments in quantum information theory over the past decade have uncovered quantum-optimal measurement methods that enable super-resolution imaging beyond the diffraction limit, and many laboratory-scale demonstrations of these methods have been performed over the past decade, culminating in the recent first on-sky demonstration of quantum-enhanced astronomical imaging. In this Review, we detail the current progress and achievements of these theoretical and experimental results, discuss the astronomical science cases motivating the development of higher resolution telescopes,  summarise the technological challenges facing the practical application of quantum imaging technologies to astronomy, and present a vision for the realization of \emph{quantum telescopes} with imaging resolutions orders of magnitude better than current observatories.
\end{abstract}


\maketitle

\section{Introduction} \label{sec:intro}
Our ability to explore and study our universe is restricted by the sensitivity and resolution of our telescopes, which are ultimately determined by the size of the telescopes. The construction of ever larger telescopes presents significant engineering challenges, particularly for optical and infra-red telescopes (spanning wavelengths from approximately 300~nm to 30,000~nm\footnote{For the purposes of this paper, ``optical" will be considered to include wavelengths from ultraviolet (approx. 300~nm - 400~nm), through the visible (approx. 400~nm - 700~nm), to the far infra-red (approx. 30,000~nm).}) where the wavelengths of the electromagnetic fields being measured place extreme requirements of the precision and mechanical stability of the telescope's optics. Recent advances in quantum information theory and quantum metrology technologies, combined under the banner of \emph{quantum imaging} hold great promise to provide solutions to these challenges and enable the construction of telescopes with orders of magnitude greater sensitivity and resolution than any telescope currently operational or under construction. These telescopes would usher in a new era of astronomical research, advancing our understanding of our universe through high-precision measurements of galaxies, star-forming regions, and stellar remnants that will shed light on phenomena from the General Theory of Relativity to nuclear matter.

\begin{figure}
\centering\includegraphics[width=0.47\textwidth]{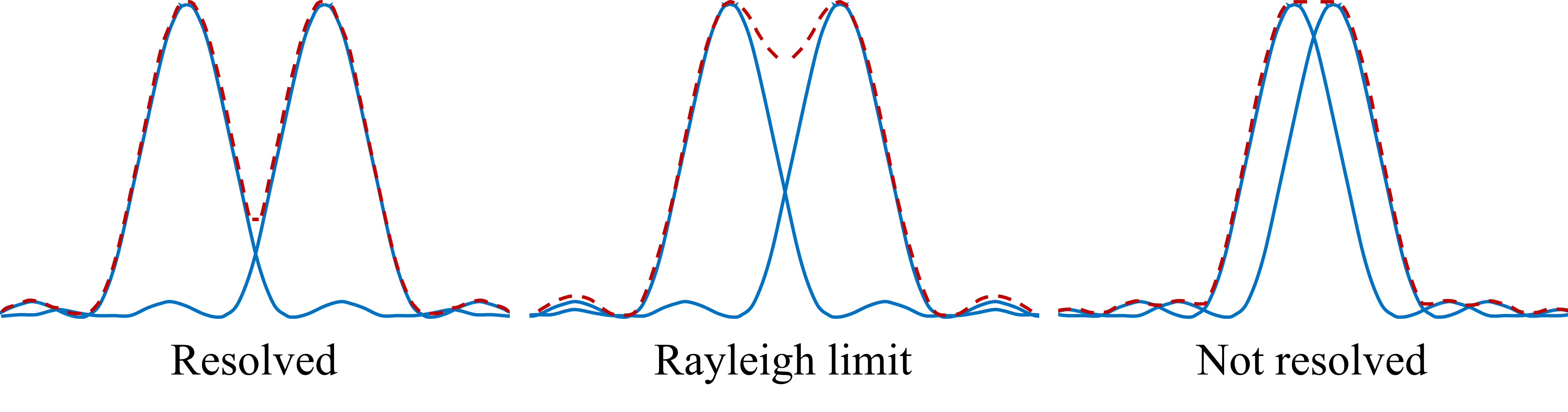}
\caption{Illustration of the Rayleigh criterion for resolution of an imaging system. The blue traces depict the intensity profile of the Airy disks resulting from two point sources through the PSF of a telescope. The red dashed trace depicts the summed intensity profile of the two Airy disks. Left-to-right: objects are resolved, objects are at the Rayleigh limit, and objects are unresolved. The Abbe and Sparrow limits are described in Section~\ref{sec:theorySum}.}
\label{fig:RayleighLimit}
\end{figure}

In 1873 Abbe \citeyearpar{abbe1873ueber} proposed a criterion to determine the limits of an imaging system to resolve two point sources. Lord Rayleigh \citeyearpar{rayleigh1879xxxi} and Sparrow \citeyearpar{sparrow1916spectroscopic} proposed similar criteria, and Rayleigh's criterion is probably the most widely known. As demonstrated in Figure~\ref{fig:RayleighLimit}, according these criteria, the sources are said to be unresolved if their images---blurred by diffraction---overlap significantly. This is the diffraction limit of the optical system. The way an imaging system blurs a point source is called the point spread function (PSF) of that imaging system. However, these criteria are heuristics, and not fundamental limits to the resolution of an imaging system. Feynman famously notes in \textit{The Feynman Lectures on Physics} that careful measurements of the exact intensity over the diffracted spots would allow better resolution \citep{feynman1966feynman}. Thus, noise in the intensity measurement (with shot noise being the dominant contributor at the fundamental level) becomes a limiting factor. In astronomy, sources are faint and so shot noise plays a significant role in limiting the achievable resolution.

The wave-particle duality of light creates two fundamental limits in imaging: the diffraction limit, and the shot noise limit. The development of quantum information theory in the latter half of the 20$^{th}$ century allowed these two limits to be considered in combination, enabling physicists to determine the measurement process that gave the best outcome. Recent developments in quantum information theory have uncovered new methods for the sub-diffraction resolution of two sources, and these methods have been demonstrated theoretically and experimentally to outperform direct (classical) imaging, and to approach the quantum limits of imaging. These methods have been shown to have benefits for both astronomy and microscopy. However, microscopy has the advantage that the object of interest can be probed with engineered light sources, or otherwise controlled by the observer to exploit the advantages of these quantum imaging methods. In astronomy, we have no such ability to control the object or the light source: the quantum state of the light field is all that is accessible for measurement. This restricts the quantum imaging methods available to astronomers.

In spite of this, quantum information theory shows that there is significantly more phase information---and thus, resolution---to be obtained even from distant observation of incoherent sources. Measurement methods that exploit this information are able to achieve \emph{super-resolution}: resolution beyond the classical diffraction limit. This Review covers the theoretical and experimental demonstrations of quantum imaging techniques that are applicable to astronomical imaging, highlights the progress to date, considers the  maturity and relative advantages of different technologies, and presents a path towards the development of practical astronomical telescopes that take advantage of quantum techniques to surpass the capabilities of current technologies. This Review also briefly considers the astronomical science that could be performed with telescopes with resolution approaching one microarcsecond (1~$\mu$as), presenting a science case for the development of such instruments. This Review is concerned not so much with quantum technologies \emph{for} astronomy---i.e. quantum technologies that will benefit or improve existing observation methods---but with the progress towards building telescopes capable of achieving a quantum leap in astronomical observing capability: a \emph{quantum telescope}. 

In Section~\ref{sec:intro} we provide a brief overview of the history, theory, and implementations of quantum imaging. Section~\ref{sec:q4astro} details the current state-of-the-art and progress of demonstrations of different quantum imaging techniques applicable to imaging of astronomical sources. Section~\ref{sec:astroGoals} considers the astronomical science that would be achievable with resolutions down to 1~$\mu$as. Section~\ref{sec:challenges} considers the current and foreseen challenges facing the translation of these quantum imaging methods from the laboratory to practical astronomy and discusses the technologies that are being developed, or need to be developed, to enable quantum telescopes to surpass the capabilities of classical telescopes across a wide range of astronomical observations. Section~\ref{sec:conclusion} concludes this Review by presenting a path towards the implementation of practical quantum imaging telescopes to provide new observing capabilities for astronomers.

\subsection{Overcoming Rayleigh's curse}
Quantum metrology studies how best to measure physical quantities. Some parameters such as phase---critical for imaging applications---do not have an associated quantum observable but quantum metrology is used to find the optimal quantum measurement from which the desired parameter can be estimated \citep{giovannetti2011advances}. Many of the current advances in quantum imaging build off the foundation of quantum detection and estimation theory developed by Helstrom in the latter half of the 20$^{th}$ century \citep{helstrom1969quantum}. Helstrom's aim was to determine, given an imaging scenario, which quantum measurement would give the optimal statistics, and thus the best inference. This optimal performance then represents the most fundamental limit of resolution that can be achieved for that scenario, according the laws of quantum mechanics.

Helstrom applied his theory to a few scenarios of incoherent imaging, including the problem of locating an incoherent point source from far-field measurements \citep{helstrom1970estimation}, and to deciding between the presence of one or two incoherent sources in the image \citep{helstrom1973resolution}. However, in all of the problems Helstrom studied, the improvements over conventional imaging were minimal, or he was unable to conceive of a method by which the necessary quantum-mechanical projection operators could be measured \citep{tsang2019resolving}.

Recently, others have applied quantum estimation theory to the problem of resolving two incoherent point sources and discovered novel far-field measurement techniques that offer resolution beyond what can be achieved with classical direct imaging.

Mankei Tsang, one of the first theorists to propose a measurement technique that promised substantial improvements over direct imaging, summarised the rapid theoretical advances and handful of experimental demonstrations of quantum imaging of incoherent point sources in 2019 \citep{tsang2019resolving}. However, the past five years have seen continued rapid growth in theoretical understanding of this field, as well as an explosion of experimental demonstrations, culminating in the recent on-sky demonstration of sub-diffraction astronomical measurements \citep{kim2025sky} using techniques derived from Tsang's proposed method. Here, we review the theoretical advances and experimental state-of-the-art in quantum imaging of incoherent sources, show that quantum imaging for astronomy is feasible, and discuss what challenges remain in realizing a practical quantum telescope with performance surpassing that of any existing telescope.

\subsubsection{Theoretical summary} \label{sec:theorySum}

Tsang's 2019 review \citep{tsang2019resolving} covered the relevant quantum information theory in detail, so we will only present a summary of theoretical results here. Other recent reviews of advances in astrophotonics and quantum technologies have briefly considered quantum imaging as applied to astronomy \citep{ellis2024astrophotonics,defienne2024advances,eisenhauer2023advances}, but have not covered the techniques and state-of-the-art in detail, or have focused only on quantum interferometers \citep{huang2026quantum}. The present review also gives greater consideration to the goals of contemporary astronomy as as driving force behind the development of quantum telescopes.

Classical imaging systems (e.g. the eye, camera, telescope, etc) are based on spatially resolved intensity measurements, typically referred to in the quantum imaging literature as \textit{direct imaging} (DI). The diffraction limit of resolution, $\theta$, of a DI system is approximately

\begin{equation}
    \theta \approx k \frac{\lambda}{D},
    \label{eq:rayleigh}
\end{equation}
where $\lambda$ is the wavelength of the light, $D$ is the diameter of the primary aperture of the imaging system, and $k$ is a constant resulting from Abbe ($k = 1$), Rayleigh ($k=1.22$) and Sparrow's ($k=0.94$) assumptions about the practical limit to which the diffracted spots could be reliably discriminated. In astronomy, the wavelength to be imaged is primarily specified by the scientific goals. For example, visible wavelengths are heavily attenuated by circumstellar and interstellar dust, greatly limiting the regions of the Milky Way and other galaxies that can be studied at these wavelengths. Infra-red wavelengths longer than around 2~$\mu$m are better able to penetrate this dust, so optical astronomy of the deep universe or small targets is increasingly concerned with observations in the near and far infra-red. 

Because of this, the quest for astronomical instruments with greater sensitivity and resolution has primarily been driven by the construction of larger apertures. At present, the largest single-aperture optical telescope is the Gran Telescopio Canarias (GTC) with a diameter of 10.4~m, and a diffraction-limited angular resolution of approximately 0.16~arcsec (0.78~$\mu$rad) at a wavelength of 8,000~nm \citep{canarias1998gran}, while the Extremely Large Telescope (ELT), currently under construction, will have a diameter of 39.3~m and a best angular resolution of 0.005~as (24~nrad) \citep{padovani2023extremely}
\footnote{For the purposes of comparison, we will ignore the effects of atmospheric seeing---the degradation of the resolution of the telescope relative to the diffraction limit due to atmospheric turbulence---and other effects that prevent a telescope from reaching its diffraction limit during normal operation.}.
The largest single-dish radio telescope---with a filled aperture---is the Five-hundred-meter Aperture Spherical Telescope (FAST), which operates at wavelengths between 0.1~m and 4.3~m \citep{nan2011five}.

To increase resolution, the light from an array of multiple apertures can be interfered in a process called \textit{aperture synthesis}, which enables the array to achieve a resolution where $D$ is now the distance between the apertures. (At the expense of reduced sensitivity due to the sparse nature of the array.) This is the principle behind radio telescope arrays such as the Square Kilometre Array (SKA) \citep{grainge2017square} and the Event Horizon Telescope (EHT) \citep{akiyama2019first}. At each separation and orientation, the interference fringe is measured to produce an output which is one component of the Fourier transform of the spatial distribution of the brightness of the observed object. The inverse Fourier transform is then taken to calculate, via the van Cittert–Zernike theorem \citep{mandel1995optical}, the visibility function to infer the brightness distribution of the source.


The EHT has a maximum baseline on the order of 10,000~km, achieving an angular resolution of 20~$\mu$as (0.1~nrad) at an operating wavelength of 1.3~mm \citep{akiyama2019first}, meaning the EHT currently holds the record for the best resolution of any telescope. Although interferometers operating at near infra-red and optical wavelengths benefit from an approximately three order-of-magnitude advantage over microwave instruments such as the EHT due to their shorter operational wavelength, optical interferometers have proved difficult to scale up, with the largest optical interferometer being the 330~m Center for High Angular Resolution Astronomy (CHARA), with a resolution of 300~$\mu$as (1.5~$\mu$rad) to 1~mas (5~$\mu$rad) from visible to near-infrared wavelengths \citep{pedretti2009imaging}.

Aperture synthesis is only possible if both the amplitude and the phase of the incoming signal are measured by each telescope. For radio frequencies, this is possible via electronics. However, for optical frequencies, the electromagnetic field cannot be measured directly and subsequently correlated in software. The received light must be propagated to a central location and interfered in real time. Because of this, optical interferometers have been limited by the low photon fluxes from optical sources\footnote{The rate of photons drops rapidly with decreasing wavelength. Many astrophysical phenomena, such as radio galaxies, pulsars and AGN, produce less flux at shorter wavelengths. This is then compounded by the greater photon energies at shorter wavelengths to produce photon rates that decline proportionally to $\lambda^\alpha$ where $\alpha > 1$.} and the sensitivity of the interference to phase noise within the array. Quantum imaging techniques offer ways around these limitations, as we shall see in Section~\ref{sec:interferometers}

Statistical treatments of resolution include parameter estimation---given a blurry and noisy image, how well can the separation of sources within the image be estimated---and hypothesis testing---how well can it be decided whether there are one or two sources and, if there are two, what is their separation?

More specifically, the resolution of an imaging system is defined by its phase sensitivity $\Delta \phi$. Conceptually, the increase in resolution achieved by an imaging system with a larger aperture can be understood from the Heisenberg uncertainty principle for Fourier Transform pairs. A larger aperture means the uncertainty about the photon's position is increased, allowing the momentum of the photon---and thus, which pixel it will arrive at in the camera---to be further constrained. Quantum measurements that are sensitive to phase enable us to maximise the measurements of the available phase information, thus maximizing the achievable resolution of the imaging system. In some cases, the available information may be no greater than what can be obtained with direct imaging, however, in many cases, quantum measurement enables us to extract additional phase information, achieving a resolution better than the diffraction limit achieved by classical optics \citep{zhou2019modern}.


 \begin{figure}
\centering\includegraphics[width=0.47\textwidth]{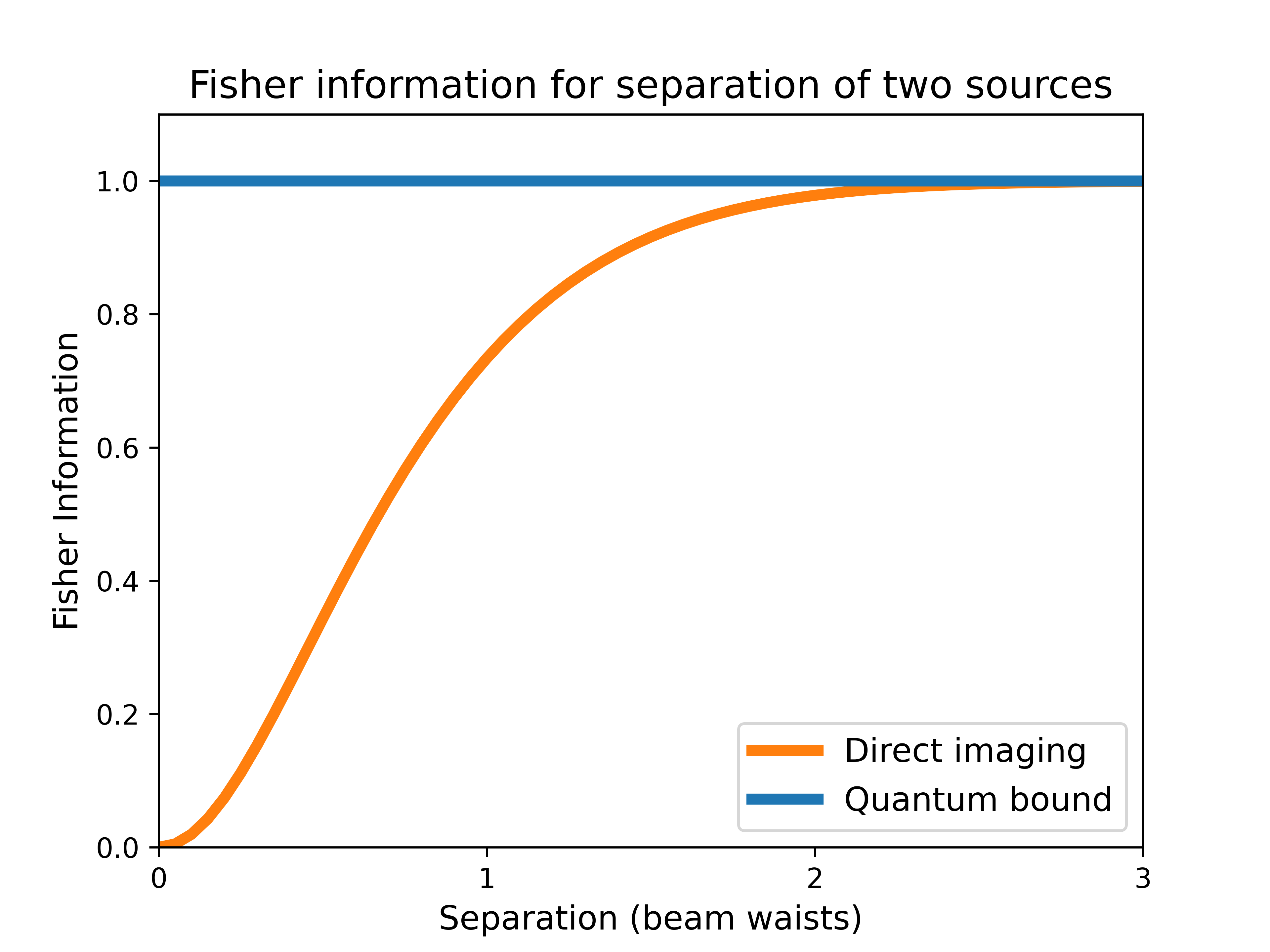}
\caption{Fisher information for the separation of two weak thermal sources according to the classical and quantum Cram\'er-Rao bounds based on \citeyearpar{tsang2019resolving}. The separation is given in multiples of the beam waist of the imaging system optics.}
\label{fig:QCRB}
\end{figure}

The theoretical analysis of the resolution of imaging systems relies on two key concepts from statistics:\\
\textbf{Fisher information:} The Fisher information (FI) quantifies the sensitivity of the data likelihood to changes in a parameter. Therefore, a larger FI implies more information about that parameter can be recovered, increasing the precision of the measurement.\\
\textbf{Cram\'er-Rao bound:} The Fisher information sets a fundamental lower bound on the variance of any unbiased estimator of a parameter (i.e. $\mathrm{Var(\hat{\theta})} = 1/F$. This is known as the Cram\'er-Rao bound. The Cram\'er-Rao bound can be violated in the case of an biased estimator. 

Helstrom proposed quantum versions of these statistical tools called the quantum Fisher information (QFI) and quantum Cram\'er-Rao bound (QCRB) \citep{helstrom1969quantum}. In 2016, Tsang, Nair and Lu applied the QCRB to derive the fundamental quantum limit for the precision of locating two weak thermal point sources \citep{tsang2016quantum}. Tsang et al. showed that, unlike the CRB which decays to zero as the separation of the sources becomes smaller than the diffraction limit, the QCRB maintains a constant value for any separation, meaning that quantum imaging protocols that are able to take advantage of these methods can exceed the diffraction limit and achieve \textit{super-resolution}. Figure~\ref{fig:QCRB} shows the Fisher information for the estimation of the separation $\theta$ between two incoherent point sources, assuming a Gaussian point-spread function. With direct imaging, the information drops to zero for $\theta \rightarrow 0$, but the Helstrom information according to the QCRB stays constant. This means that, if the optimal quantum measurement can be implemented, an imaging system can significantly surpass the diffraction limit.

Notably, Tsang et al. propose one such imaging method that reaches the maximum of the QCRB: spatial-mode demultiplexing (SPADE) \citep{tsang2016quantum}, which is described further in Section~\ref{sec:spade}.

\subsubsection{Circumventing the Rayleigh limit} 

Quantum or ``quantum-inspired" imaging techniques have found great success in microscopy and lithography because of the ability to engineer and control the light source. Such methods include: stochastic optical reconstruction microscopy \citep{rust2006sub,betzig2006imaging,fernandez2008fluorescent}, a fluorescence imaging technique that exploits the the physical structure of the object being studied; non-classical states of light where $N$ photons are entangled at a time, enabling features with a minimum size of $\lambda/(2N)$ to be written in lithography \citep{boto2000quantum,d2001two} or enhanced intensity in photodetection \citep{lloyd2008enhanced}; ghost imaging \citep{lemos2014quantum}, where entangelment is used to to extract imaging information without detecting the photons that passed through the imaged object; and $N00N$ states---a superposition of $N$ photons across two arms of an interferometer \citep{dowling2008quantum}---have been demonstrated to enable phase super-resolution in quantum interferometers \citep{slussarenko2017unconditional}. A more complete review of current quantum imaging techniques was recently provided by Defienne et al. \citeyearpar{defienne2024advances}.

However, all of the above quantum imaging methods rely on the ability to engineer the light source, or otherwise probe the object of interest with controlled states of light. However, in astronomy, the objects are remote and the quantum state of the light field is all that is accessible to be measured. Following Tsang et al., in 2017 Sidhu and Kok found multiple optimal estimators for the measurement of the angle of separation, $\theta$, \citep{sidhu2017quantum}. Also in 2017, {\v{R}}eha{\v{c}}ek showed that, while for equally bright sources the QCRB is independent of their separation, for the general case of unequal sources, the amount of information available about the separation falls to zero, but there is still a quadratic improvement in comparison with direct imaging \citep{vrehavcek2017multiparameter,vrehavcek2018optimal}. For a recent review on the QFI and the fundamentals of quantum estimation theory, see Sidhu and Kok \citeyearpar{sidhu2020geometric}. These results led to a great increase in effort by theorists to apply quantum information methods to imaging problems in astronomy. Following the first small-scale proof-of-principle laboratory demonstrations (detailed in Section~\ref{sec:q4astro}), the ready availability of sources and suitable single photon detectors has lead to a rapid increase in capabilities of experimental demonstrations of quantum imaging of incoherent sources.

\section{Quantum imaging for astronomy} \label{sec:q4astro}

\subsection{Hanbury Brown-Twiss intensity interferometer} \label{sec:HBT}

The first astronomical measurement technique to exploit the quantum nature of photons for advantage over conventional techniques was the \textit{intensity interferometer} pioneered by Robert Hanbury Brown and Richard Twiss \citep{hanbury1954lnew,hanbury1956correlation,hanbury1956sirius,hanbury1968stellar,hanbury1974angular}. The Hanbury Brown-Twiss (HBT) effect refers to correlation and anti-correlation effects of a beam of particles in the intensities received by two detectors. 

Intensity interferometers exploit the second-order correlations, $g^{(2)}$, of light instead of the more familiar first-order correlations, $g^{(1)}$, used by an amplitude (Michelson) interferometer. The second-order correlation function of light $g^{(2)}$ with the time-variable intensity $I(t)$ is given by

\begin{equation}
    g^{(2)}(\tau) = \frac{\langle I(t)I(t+\tau)\rangle}{\langle I(t)\rangle^2},
    \label{eq:HBTg2}
\end{equation}
where $\tau$ is the correlation time delay, $t$ is time, and $\langle \rangle$ denotes long-term averaging \citep{dravins2008toward}.

For thermal emission, the relation between the first-order coherence function and the second-order function is simply given by $g^{(2)} = |g^{(1)}|^2 - 1$. The modulus of the visibility $g^{(1)}$ can be directly determined from intensity-correlation measurements.

Hanbury Brown and Twiss initially applied this concept to radio astronomy, and this method was used to measure the angular size of radio stars \citep{hanbury1956correlation,jennison1956v}. They suggested the technique should work for visible light as well, and demonstrated the the effect in the laboratory using a mercury-vapour lamp as the source \citep{hanbury1956correlation} before constructing larger instruments to measure the angular size of stars, starting with Sirius ($\alpha$ Canus Majoris A) \citep{hanbury1956sirius}, and then producing a catalogue of the measurements of a total of 32 stars \citep{hanbury1968stellar,hanbury1974angular}. At the time, Hanbury Brown's results were met with skepticism, with many believing that the HBT effect did not make physical sense. The dispute sparked the development of the field of quantum optics \citep{fano1961quantum,glauber1963photon,aspect2020hanburry}. Most notably, to give a fully consistent description of the results of the HBT experiment, Glauber developed the quantum optics formalism still used today \citep{glauber1963photon,aspect2020hanburry} for which he shared the 2005 Nobel Prize in Physics.

\subsubsection{The Narrabri Stellar Intensity Interferometer} \label{sec:pastHBT}

\begin{figure}
\centering\includegraphics[width=0.47\textwidth]{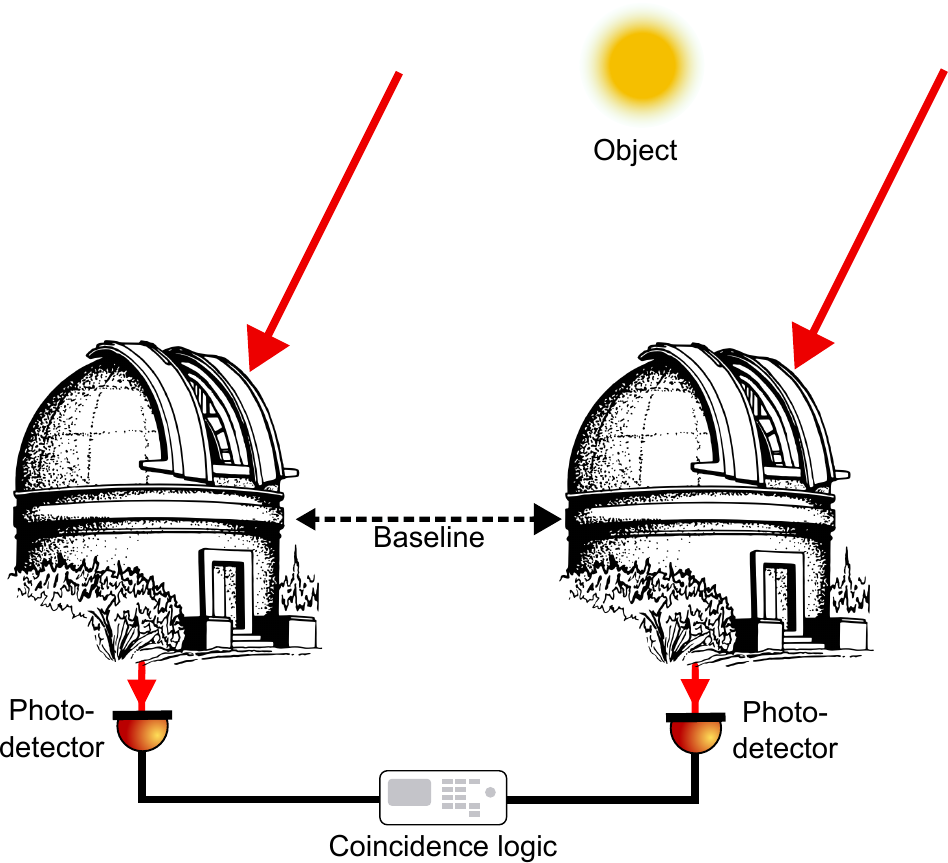}
\caption{Generalised schematic of an intensity interferometer based on the Hanbury~Brown-Twiss (HBT) effect.}
\label{fig:HBT}
\end{figure}

A stellar intensity interferometer based on the HBT effect is built from two light detectors---``light" here referring to any part of the EM spectrum---separated by some baseline. Intensity measurements of the source are correlated to measure the interferometric visibilities, which can then be used to calculate the angular diameter and limb darkening coefficients of stars. The advantage of intensity interferometers is that only the intensity measured at each detector needs to be transmitted to the correlator, rather than the amplitude and phase of the signal.\footnote{Interferometers that use phase information are variously referred to as amplitude interferometers, phase interferometers, or Michelson interferometers in the literature. Here, we will refer to them as \emph{amplitude interferometers}.} This lack of phase information makes intensity interferometers relatively easy to implement, and insensitive to phase noise during transmission of the signal or resulting from atmospheric turbulence. However, the lack of phase information means that intensity interferometers cannot produce aperture synthesis images because the visibility phase information is not preserved.

Hanbury~Brown's interferometer at Narrabri in New South Wales, Australia, summarised in Figure~\ref{fig:HBT}, comprised two 6.7~m diameter reflectors with photomultiplier tubes at the focus. The two reflectors ran on a circular track that allowed the baseline to be varied between  10~m and 188~m and maintain the axis of the baseline normal to the direction of the star. Fluctuations in the photomultiplier anode currents over a bandwidth from 10-100~MHz were carried to a central correlator via cables \citep{hanbury1968stellar,hanbury1974angular}.

The $g^{(2)}$ measurement that the HBT effect is dependent on relies on post-selection of two-photon coincidence events, which are much rarer than one-photon events, greatly reducing the single-to-noise ratio (SNR) of intensity interferometers. For example, Davis and Tango demonstrated a prototype amplitude interferometer in 1986 by measuring the angular size of Sirius with only 2\% of the observation time required by Hanbury~Brown's intensity inteferometer \citep{davis1986new}. 
Amplitude interferometry techniques that measure $g^{(1)}$, relying on the much more abundant one-photon events, have distinct advantages in terms of SNR, and thus limiting magnitude, over intensity interferometers. 
Because of this, intensity interferometers have been obsolete for decades. However, advances in quantum optics, quantum technologies, and electronics mean that modern intensity interferometers are of interest for specific imaging applications in astronomy.

\subsubsection{Future of intensity interferometers} \label{sec:futureHBT}

The relative ease of implementation of intensity interferometers compared with amplitude interferometers or large single-aperture telescopes means that intensity interferometers are still considered by many in the astronomical instrumentation community to have merit.

While a two-element intensity interferometer is unable to image a source---that is, distinguish whether a source is asymmetric and thus derive model-independent observations---this can be circumvented by the addition of further baselines that permit a full reconstruction of the source. Because intensity interferometers can analyze the electronically recorded signal, they can perform lossless comparison of many baselines, something optical amplitude interferometers cannot do \citep{dravins2008toward}.
Using intensity correlations of more than two photons would allow HBT interferometers to approach closer to the CRB and achieve super-resolution \citep{oppel2012superresolving,pearce2015precision}, but such events would be exceptionally rare and require extremely large collectors.

While large collecting areas are required to provide the intensity interferometer with sufficient flux---Hanbury~Brown's Narrabri instrument had 6.5~m diameter collectors, larger than any telescope at the time---the signal is largely independent of the spectral pass band, meaning the SNR can be improved by observing over wider bands \citep{brown1974intensity,davis1975proposed,dravins2008toward,dravins2012stellar,trippe2014optical,tuthill2024stellar}. However, this requires faster electronics, with timing accuracies of 100~ps or better, requiring the mirror to have a mechanical accuracy better than millimetres, and beginning to require the inclusion of adaptive optics (AO) systems to suppress path-length differences imposed by atmospheric turbulence \citep{dravins2008toward,tuthill2024stellar}. 

Various proposals have been made to add intensity interferometer backends to air Cherenkov experiments---in particular, the Cherenkov Telescope Array (CTA)---in order to take advantage of the existing infrastructure of the array (i.e. large mirrors of low optical quality with high-speed single photon detectors) \citep{dravins2008toward,acciari2020optical,bojer2022quantitative,tuthill2024stellar,lebohec2006optical,nunez2012high}. Some groups have investigated linking large (8~m class) optical telescopes in order take advantage of their size and adaptive optics systems \citep{baril2010ohana,tuthill2024stellar} or using sub-apertures of future large telescopes \citep{barbieri2007astronomical,dravins2005quanteye}, while others have proposed constructing dedicated intensity interferometers \citep{horch2013intensity,tuthill2024stellar,trippe2014optical}.

If the latter two options are implemented, the resulting intensity interferometer is predicted to have a sensitivity improvement over the NSII of a factor of 24,000, equivalent to 11 stellar magnitudes \citep{trippe2014optical,horch2013intensity,tuthill2024stellar}. Aspects of the various proposals have already been simulated or tested \citep{dravins2015long,guerin2018spatial,acciari2020optical,abeysekara2020demonstration,zmija2024first,baril2010ohana}.

Bojer and colleagues recently applied the QFI analysis to compare intensity and amplitude interferometers \citep{bojer2022quantitative}. They concluded that intensity interferometers with baselines of 2~km to 10~km offer practical advantages for astronomy in spite of their much poorer SNR. However, this comparison was done assuming that amplitude interferometer telescopes would not get significantly larger than the ELT (39.3~m), VLTI (130~m), or CHARA (330~m). As we shall see in Section~\ref{sec:interferometers}, quantum imaging technologies promise solutions to allow optical amplitude interferometers to exceed these scales.

Recently, Stankus, Nomerotski, Slosar, and Vintskevich proposed an interferometer architecture combining the HBT effect with the Hong-Ou-Mandel (HOM) effect to create a two-photon amplitude interferometer \citep{nomerotski2020quantum,stankus2022two,chen2023astrometry}. (This architecture is referred to as the SNSV scheme.) The SNSV scheme comprises two observing stations collecting photons from two sky sources. At each station, the photons from the two sources are interfered at a beam splitter. Correlations of photon detections at the four photodetectors (two detectors at each observing station) are sensitive to the relative phase difference of the photons from the two sources. The angle between the two sources can then be derived from the correlation statistics.

The SNSV method retains a key advantage of the intensity interferometer, that high-precision optical links between the observing stations are not required, and has the advantage over intensity interferometers that phase information is retained, allowing the complex visibility to be measured, permitting full imaging of the source. Like the proposed intensity interferometers above, an SNSV interferometer can perform the measurement over many spectral bins at once, increasing SNR. However, the SNSV method retains the disadvantage of deriving its measurements from the relatively rare two-photon states, and requires higher precision optics at the observing stations.
The SNSV method has been successfully demonstrated in the laboratory \citep{crawford2023towards}.

While astronomical interferometers based on the HBT effect have received little attention in the decades since Hanbury~Brown's pioneering work with the NSII, the improvements achievable with modern technologies mean that the benefits of two-photon interferometers to astronomy are being revisited. It is likely that a two-photon interferometer, probably an intensity interferometer, either piggybacking on other telescope infrastructure or as a bespoke instrument, will be used to complement single-photon amplitude interferometers in the future. As will be seen in Section~\ref{sec:interferometers}, developments in quantum technologies are likely to enable a rapid increase in the baseline distances of single-photon interferometers in the near future.

\begin{figure}
\centering\includegraphics[width=0.47\textwidth]{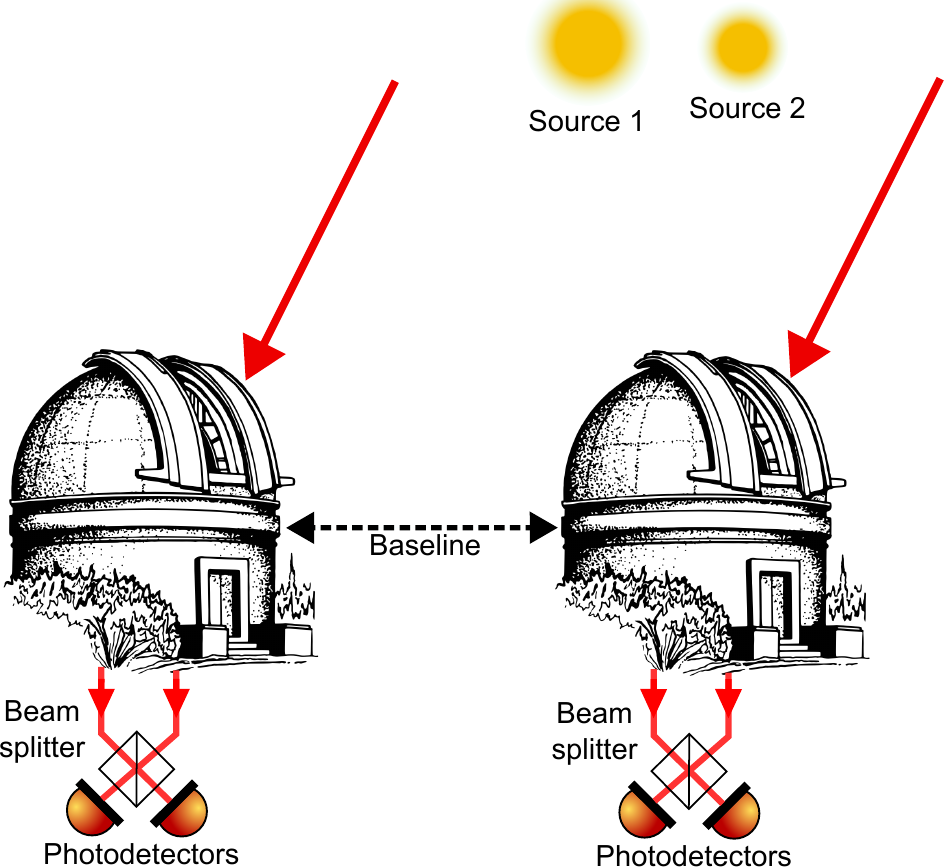}
\caption{Generalised schematic of a Stankus-Nomerotski-Slosar-Vintskevich two-photon amplitude interferometer. Two-photon states from two optical sources are interfered at the beam splitter in each observing station. Correlations between the measurements at the two stations determine the source visibility. Measurement data can be transmitted between the stations electronically.}
\label{fig:SNSV}
\end{figure}

\subsection{Spatial mode demultiplexing (SPADE)} \label{sec:spade}

\begin{figure}
\centering\includegraphics[width=0.4\textwidth]{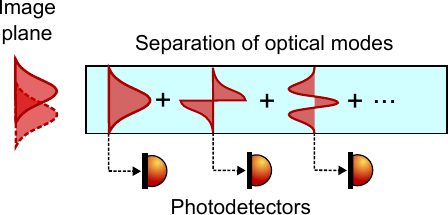}
\caption{Generalised schematic of a SPADE imaging system. Each mode is separated and coupled to a separate photon-counting detector.}
\label{fig:SPADE}
\end{figure}

As already noted in Section~\ref{sec:theorySum}, Tsang, Nair, and Lu applied the QCRB to determine the fundamental limit of precision in locating two weak thermal point sources and, importantly, showed the QCRB maintains a constant value for any separation \citep{tsang2016quantum,tsang2017subdiffraction,tsang2019semiparametric}, as shown in Figure~\ref{fig:QCRB}. In that same work, Tsang and colleagues proposed a general method to realise their quantum-limited measurement, shown in Figure~\ref{fig:SPADE}, which they dubbed SPADE.

Instead of measuring the position of each photon at the image plane, as in direct imaging, SPADE works by discriminating the Hermite-Gaussian spatial modes of the image plane field. The modes are decomposed and the flux in each mode is directed to a photon-counting detector to measure the weight of that mode.

Tsang et~al. provide an intuitive semi-classical explanation for the operation of SPADE: that it takes advantage of the sensitivity of the mode coupling to the offset of the wavefunctions from the centroid \citep{tsang2016quantum,tsang2019resolving}. Similar to how a Fourier transform provides a decomposition of a real function into a superposition of sine and cosine terms, the SPADE measurement decomposes the overlapping PSFs into the basis of Hermite functions. By measuring the flux in the higher-order modes, SPADE extracts additional phase information, compared with direct imaging, providing super-resolution and beating the diffraction limit.

Tsang also notes that direct measurement of the flux in each mode is a more optimal measurement because homodyne or heterodyne detection would suffer from excess vacuum noise introduced at the beam splitter \citep{tsang2016quantum,tsang2019resolving}.

As we will see in Sections~\ref{sec:SLIVER} to \ref{sec:SPADEsum}, there are several methods for discriminating spatial modes and measuring the flux in a chosen mode. However, the method that most closely resembled Tsang's original proposal, and may hold most promise for a practical SPADE imaging method of astronomy, uses a multi-planar light converter (MPLC) to perform the mode decomposition. As will be discussed further in Section~\ref{sec:MPLC}, MPLC's take in a multi-mode light field and direct each spatial mode out of a separate output. 

Boucher and colleagues were the first to demonstrate the benefits of MPLCs for SPADE imaging \citep{boucher2020spatial}. Although they did not specifically image a source analogous to an astronomical source, their analysis suggested that SPADE imaging with an MPLC could discriminate sources of equal intensity separated by as little as 5\% of the waist of the sources \citep{boucher2020spatial}. (i.e. the PSFs produced by the sources are 95\% overlapped.)

Since this work, others have experimentally demonstrated various aspects of SPADE imaging with MPLCs. Tan and colleagues calibrated the response of an MPLC and simulated its response to a pair of sub-diffraction sources of equal intensity, again showing that SPADE imaging can achieve super-resolution imaging \citep{tan2023quantum}. Importantly, Tan et~al. showed that imperfections in the MPLC allowed them to resolve sign ambiguities in the response of the MPLC and localise the sources as well as measure their separation.

Santamaria and colleagues tested the response of an MPLC to sources of unequal brightness---a far more realistic, and thus important, scenario for astronomical imaging. They showed that by measuring the flux in the HG01 and HG10 modes output by the MPLC they could resolve sources of equal brightness separated by only 2.4\% of the beam waist and, for a fixed separation of 66\% of the beam waist, they could determine the presence of a secondary source 10$^5$ times weaker than the primary source \citep{santamaria2023spatial}. Satamaria et~al. achieved this measurement with photodetectors that were not sensitive to single photons, simply measuring the photocurrent produced by the HG01 and HG10 modes. 

Rouvi{\`e}re and colleagues performed super-resolution SPADE imaging of sources of equal intensity in both a high-flux regime using conventional photodetectors (Thorlabs PDA50B2, 10$^{13}$ detected photons) and low-flux regime using single-photon detectors (IDQuantique ID230, 3500 detected photons), and demonstrated a resolution five orders of magnitude better than the diffraction limit. \citep{rouviere2024ultra}. 

Santamaria and colleagues have also recently repeated their experiment, imaging two sources of unequal intensity, with single-photon detectors \citep{santamaria2024single}. They achieve residual errors only 2-4 times the theoretical limit of SPADE imaging, and these errors are dominated by imperfections in the mechanical translation stages in the laboratory source, rather than the MPLC itself, although the MPLC does suffer from crosstalk. By comparing their SPADE measurements to direct imaging, they show that SPADE yields an SNR improvement of $\epsilon^{-1/2}$ over direct imaging, which shows promise for the application of SPADE to sources of extremely different intensity, such as imaging exoplanets. Santamaria and colleagues have also recently demonstrated super-resolved spectroscopy using SPADE \citep{amato2024single,huang2023ultimate}. Other experiments have demonstrated the potential of SPADE to create a super-resolution coronagraph by discarding the modes primarily coupling the star \citep{deshler2025experimental}.

Santamaria \textit{et~al.} were the first team to experimentally investigate estimation of both the separation, $d$, and relative intensity, $\epsilon$, of sources using SPADE. (Rouvi{\`e}re and Boucher used sources of equal intensity, and so only estimated separation.) However, Santamaria \textit{et~al.} only consider estimating these parameters separately, and do not demonstrate estimation of both parameters simultaneously. Wallis \textit{et al.} \citeyearpar{wallis2025spatial} perform simultaneous estimation of the sub-diffraction separation and relative brightness of two sources using maximum likelihood estimation while Gozzard \textit{et al.} \citeyearpar{gozzard2025super} perform the same analysis using machine learning. Both of these studies fail to reach the same performance as Santamaria \textit{et al.} due to crosstalk between modes in the demultiplexer.

Despite these demonstrations, extending SPADE to measurements of more than two sources (i.e. conducting 2D imaging of spatially extended sources of complex shapes) remains a challenge. There is no fundamental reason preventing QCRB-limited 2D imaging, however, the complexity of the parameter estimation becomes prohibitive. Methods that could potentially overcomes this include neural networks \citep{frank2023passive,pushkina2021superresolution,gozzard2025super,buonaiuto2025machine,sajia2025breaking,kuusela2025optical}, iteratively updating both the form of the spatial imaging modes and the estimate of the source distribution \citep{matlin2022imaging,desai2025multiple,choi2025super}, and moment-based super-resolution measurements \citep{sorelli2021moment}. Quantum computing methods also provide a potential solution \citep{mokeev2025enhancing}.

SPADE can image in three dimensions, that is, resolve the separation of sources in depth as well as angular separation \citep{zhou2019quantum}. This is potentially an extremely useful capability for mapping the distribution of objects within star and galaxy clusters, but further exacerbates the challenge of effective parameter estimation of extended or complex sources.

Santamaria's recent work \citep{santamaria2024single} also gives the most detailed analysis to date of the effects of noise and crosstalk in SPADE imaging with an MPLC. Imperfections and crosstalk are always going to prevent SPADE from reaching its ultimate quantum limits, even to the extent of performing more poorly than direct imaging under some conditions, however, analyses of realistic crosstalk and source separation indicate SPADE is superior over many orders of magnitude of source separation \citep{gessner2020superresolution,linowski2023application,schlichtholz2024practical}. Most worryingly for applying SPADE to astronomy, it has been shown that any nonzero crosstalk leads to a breakdown of super-resolution when the number $N$ of detected photons is large.

Most laboratory demonstrations of SPADE have dealt with the sub-diffraction scenario, with few demonstrations of measurements at larger separations. This is likely because the ability to resolve sub-diffraction separations is a convincing demonstration of the veracity of the SPADE method. However, demonstration of effective measurement of the source distribution at sub-diffraction \textit{and} super-diffraction separations is required for practical astronomy.
Nair and Tsang show that SPADE approaches the QCRB over greater ranges of separation as the number of measured modes is increased \citep{nair2016far}.

The SPADE imaging method is extremely sensitive to misalignment between the centroid of the object and the central axis of the imaging receiver and misalignment beyond a few percent quickly degrades the imaging mean square error (MSE) \citep{tsang2016quantum,tsang2019resolving,grace2020approaching,grace2022identifying}. However, determination of the optical centroid of an unresolved source is easily achieved with a high degree of precision using direct imaging.
Grace and colleagues theoretically investigated using conventional direct imaging to guide the centroid determination for SPADE \citep{grace2020approaching}. They concluded that a two-stage imaging system that switched from SPADE to direct imaging was better for imaging super-diffraction targets because of the degradation SPADE suffers from centroid misalignment. However, using a direct imaging system to guide a SPADE imaging system will enable the SPADE system to be aligned to the optical centroid of the source with an error of a fraction of a percent, so is not going to be a limiting source of measurement error compared with other noise sources in realizable SPADE systems. Grace and Guha also proposed and analyzed a method they call TRISPADE, which measures the outputs of three spatially demultiplexed modes and find that this method is robust to centroid misalignment \citep{grace2022identifying}. Centroid estimation has no equivalent to the diffraction limit. The centroid may be measured using direct imaging, and errors reduce by the factor $1/n$ for data points with $n$ much greater than 1.

Until recently, most leading SPADE experiments have used multi-planar light converters (see Section~\ref{sec:MPLC}) to perform the spatial mode demultiplexing. Photonic lanterns are an alternative technology that were developed for applications in astronomical measurement. This maturity in astronomical applications recently lead to the first on-sky demonstration of sub-diffraction measurements using SPADE \citep{kim2025sky}, achieving a photocenter precision of 50~$\mu$as, and highlighting the outstanding potential of SPADE for astronomy.

\subsubsection{Spatial mode demultiplexer technologies} \label{sec:MPLC}

Multi-planar light converters \citep{labroille2014efficient,fontaine2017design,fontaine2018multi,zhang2023multi} are currently the most promising technology for an effective implementation of SPADE in astronomy. This is highlighted by the number of experimental demonstrations of SPADE using MPLCs \citep{boucher2020spatial,tan2023quantum,santamaria2023spatial,rouviere2024ultra,santamaria2024single,wallis2025spatial,gozzard2025super}.

MPLCs uses a series of phase masks to convert one orthogonal set of beams into another orthogonal set through unitary transformation \citep{labroille2014efficient,fontaine2017design,fontaine2018multi,zhang2023multi}. They were initially developed for mode-division multiplexing in optical telecommunications. For the purposes of SPADE, MPLCs can be used to receive a multi-mode light field and sort the flux in each mode into a separate single-mode fibre output.

The essence of SPADE imaging is the measurement of signal intensity following some form of mode discrimination. Sections~\ref{sec:SLIVER} to \ref{sec:SPADEsum} present other possible implementations of SPADE but, as we shall see, all of them have significant disadvantages over the use of MPLCs.

All of the experimental demonstrations of SPADE imaging with MPLCs have been performed at a wavelength of 1550~nm \citep{boucher2020spatial,tan2023quantum,santamaria2023spatial,rouviere2024ultra,santamaria2024single}. This is because all of these experiments have relied on commercially available MPLCs (namely, the Cailabs PROTEUS-C) which were designed for mode-division multiplexing of signals in the telecommunications C-band. Demonstrations of MPLCs operating at wavelengths of 2~$\mu$m and longer will be of interest to the astronomical instrumentation community. Maximizing the optical bandwidth that can be handled by a single MPLC will also be critical for increasing the sensitivity of a SPADE-based telescope to practical levels.

A technology similar in functionality to the MPLC that has recently found application in SPADE is the \emph{photonic lantern} \citep{ellis2024astrophotonics,birks2015photonic,leon2013photonic,leon2017photonic,kim2025sky}. Photonic lanterns also take in a multi-mode light field and couple the field into multiple single-mode outputs. However, unlike the MPLC, a basic photonic lantern does not direct a specific HG mode to a specific output. HG modes entering the photonic lantern are mapped into ``lantern modes", with each lantern mode comprising some non-deterministic weighted sum of the intensities in each received HG mode. While this mapping is non-deterministic, it is fixed at the time of manufacture of the photonic lantern, and is stable except in cases of extreme heat or stress applied to the lantern. This means that the mode mapping from input to output can be calibrated for each photonic lantern. 

Photonic lanterns can be scaled to large numbers of modes more easily than MPLCs, potentially giving photonic lanterns an edge over MPLCs for implementations of SPADE with a large number of measured modes. Mode-selective photonic lanterns, more analogous to the functionality of an MPLC, do exist, but they are more difficult to make and less reliable in terms of mode crosstalk and isolation \citep{leon2014mode}.

No direct comparison of the super-resolution imaging performance between MPLC and photonic lantern methods has been performed but, thanks to the use of relatively mature photonic lantern technology in astronomy for spectroscopy, a photonic lantern was recently used to perform the first on-sky demonstration of sub-diffraction SPADE imaging using the 8.2~m Subaru Telescope \citep{kim2025sky}. The authors reported an impressive photocenter precision of 50~$\mu$as, but did not report the angular resolution of the system.

Photonic lanterns are also able to act as simultaneous wavefront sensors and imaging systems, allowing them to derive control signals for the telescope's adaptive optics systems with significant advantages over conventional wavefront sensors, such as Shack-Hartman wavefront sensors, by virtue of being located at the image plane, rather than the pupil plane of the telescope \citep{norris2024photonic,eikenberry2023photonic}.

\subsubsection{Binary SPADE} \label{sec:biSPADE}

\begin{figure}
\centering\includegraphics[width=0.4\textwidth]{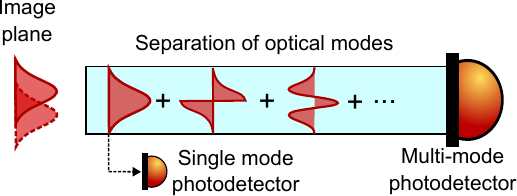}
\caption{Generalised schematic of a binary SPADE imaging system. Light in a desired mode is tapped-off and directed to a photon-counting detector, while light in all other modes is directed to a second photon-counting detector.}
\label{fig:biSPADE}
\end{figure}

In Tsang's original SPADE proposal \citep{tsang2016quantum} the authors also proposed a version they dubbed \emph{binary} SPADE. This is a simplification of the SPADE design where only one desired mode is coupled out separately to one photodetector, while the light in all other modes is detected by a second photodetector as depicted in Figure~\ref{fig:biSPADE}. 

Binary SPADE takes advantage of the fact that direct imaging has no disadvantages over SPADE for source separations larger than the diffraction limit. Below the diffraction limit, only low-order HG modes (i.e. HG$_{01}$ and HG$_{10}$) have significant flux coupled into them. However, it has been shown that binary SPADE is quantum-optimal for all values of separation \citep{lu2018quantum}, potentially making binary SPADE an extremely useful method for specific astronomy applications such as measurements of the angular size of stars or the detection and parameter estimation of two-source targets such as exo-planets. For the purposes of source discrimination or stellar size measurements, the chosen single mode would be the HG00 mode, while the combined flux from all other modes is detected together. 

Wallis \textit{et al.} demonstrated binary SPADE in the laboratory using a simple double-clad fibre coupler \citep{de2015double} and two photodetectors to provide super-resolved discrimination between one and two sources \citep{wallis2025super}, showing the potential application of binary SPADE to specific astronomical measurement cases.

\subsection{SLIVER} \label{sec:SLIVER}

\begin{figure}
\centering\includegraphics[width=0.47\textwidth]{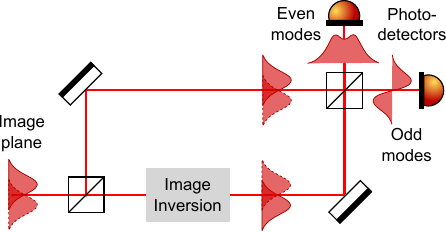}
\caption{Generalised schematic of an image-inversion interferometer for SLIVER. The interference of the original image with the inverted image causes even modes to be coupled to one port of the beamsplitter, while the odd modes are coupled to the other port.}
\label{fig:sliver}
\end{figure}

Implementation of SPADE requires that different spatial modes be coupled into physically separate channels before detection. In 2016, Nair and Tsang proposed the superlocalization via image-inversion interferometry (SLIVER) method that can, in principle, achieve quantum-limited Fisher information for $\theta \rightarrow 0$.

The setup for SLIVER is shown in Figure~\ref{fig:sliver}. The image light is directed into the input of a Mach-Zehnder interferometer with one of the arms containing optics to invert the image. This inversion can be implemented with mirrors, lenses, or a Dove prism. The interference of the original and inverted images at the recombining beamsplitter results in all of the even modes on the image plane coupling out of one port of the interferometer, while all of the odd modes are coupled out of the other port, allowing the even fundamental mode, $\phi(x)$, to be measured.

Tang and colleagues demonstrated a proof-of-concept SLIVER in 2016, but did not reach the quantum limit \citep{tang2016fault} though later experiments have demonstrated the potential of this method to reach the quantum limit \citep{larson2019common,aiello2025sub,katamadze2025breaking}. Importantly for astronomical imaging, image-inversion interferometers have long been known to be less susceptible to atmospheric turbulence than direct imaging \citep{roddier1988interferometric}. This combination of resilience to turbulence and quantum-limited resolution make SLIVER a technique with exciting potential for astronomy. 

Nair and Tsang proposed a variant dubbed pix-SLIVER which introduces an additional reflection in one arm of the Mach-Zehnder interferometer to separate the image-plane field into its symmetric and antisymmetric components, and replace the two detectors with detector arrays at the outputs of the beam splitter \citep{nair2016far}. Lu and colleagues showed that the pix-SLIVER method is near-optimal for sub-diffraction separations \citep{lu2018quantum}. Nair and Tsang show that pix-SLIVER approaches the QCRB over greater ranges of source separation as the number of pixels is increased \citep{nair2016far}. (Similar to how SPADE approaches the QCRB over greater ranges of separations as the number of measured modes is increased.)

A single SLIVER device is only capable of estimating the separation of two sources in one dimension. To extend the SLIVER method to more general imaging scenarios, it would be necessary to construct multiple inversion interferometers, each measuring a different mode, in the manner of a fractional Fourier transform \citep{xue2001beam,martin2017basis,hassett2018sub,zhou2019quantum}. The statistical performance of such a setup has not been investigated, but the overall performance of a multi-mode SLIVER would likely be poor compared to methods such as the MPLC because of the inefficient use of available photons as all modes would be directed to each of the interferometers. 

Most theoretical and experimental investigations of SLIVER focused on assessing the effectiveness of the method in imaging two sources of equal brightness. Recently, Zhang and colleagues extended the theoretical analysis to two sources of unequal brightness \citep{zhang2024superresolution}; a far more realistic scenario in astronomical observations.

\subsection{SPLICE}

\begin{figure}
\centering\includegraphics[width=0.47\textwidth]{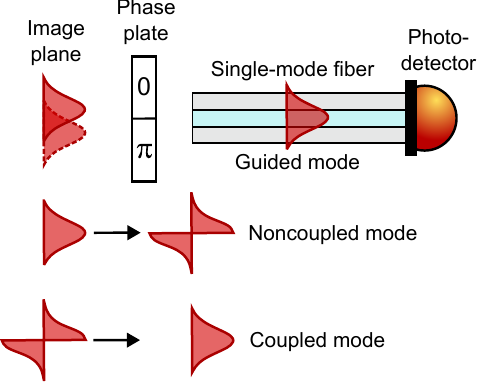}
\caption{Generalised schematic of the SPLICE imaging method. A phase plate is used to introduce a $\pi$ phase shift to one half of the image plane. The odd mode is converted by the phase plate into the fundamental single mode of the fibre and propagates to the detector.}
\label{fig:splice}
\end{figure}

In 2017, Tham and colleagues proposed and experimentally demonstrated a method they called super-resolving position localization by inversion of coherence about an edge (SPLICE) \citep{tham2017beating}. As shown in Figure~\ref{fig:splice}, the SPLICE method uses a phase plate to introduce a $\pi$ phase shift to half of the image plane. This phase shift converts odd modes to even modes, allowing them to couple efficiently into single-mode fibre and propagate to the detector. Orthogonal modes are rejected by the fibre. 

Despite the imperfect match between the converted odd mode and the mode selection of the fibre, multiple groups have demonstrated super-resolution with this technique \citep{tham2017beating,wadood2024experimental}. Other methods to convert a chosen higher order mode for efficient fibre coupling, such as long period gratings \citep{amorim2025spatial}, offer ways to improve the practicality of this technique.

While the SPLICE method is elegant in its simplicity---it essentially modifies the desired incoming mode and uses a single-mode fibre for effective spatial mode selection---the rejection of photons not in the desired mode is inefficient. The SPLICE method may have some applications in model-dependent imaging such as discriminating the presence of an exoplanet or other companion object.

\subsection{Digital holography}

\begin{figure}
\centering\includegraphics[width=0.35\textwidth]{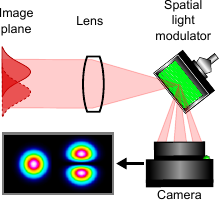}
\caption{Generalised schematic of imaging using a hologram to perform mode sorting. A lens images a field onto a hologram placed at the image plane which projects the incoming light into its constituent modes. The diffraction spectrum is mapped onto a camera.}
\label{fig:hologram}
\end{figure}

Holograms can be used to perform spatial matched filtering. In this form of imaging, the hologram is designed such that the intensities of the diffracted points in the far field are proportional to the mode spectrum \citep{goodman2005introduction,forbes2016creation}. Pa{\'u}r and colleagues demonstrated this technique in 2016, using a spatial light modulator as shown in Figure~\ref{fig:hologram} to create the necessary hologram to project the mode spectrum onto an EMCCD, with their MSEs reaching close to the QCRB \citep{paur2016achieving}.

While Pa{\'u}r used a spatial light modulator to create the hologram, the hologram does not need to change, so can be fixed at manufacture. For astronomy purposes, implementing this method could be as simple is placing a fixed holographic plate in front of an EMCCD \citep{zhou2023approaching}, which are already widely used in astronomy.

\subsection{Homodyne and heterodyne detection}

\begin{figure}
\centering\includegraphics[width=0.3\textwidth]{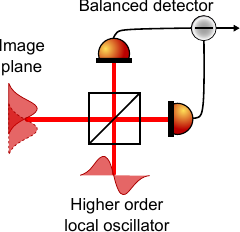}
\caption{Generalised schematic of imaging using balanced homodyne or heterodyne detection. The image light is interfered with a local oscillator in a higher-order mode and detected. The resulting signal is used to infer the source separation.}
\label{fig:BHD}
\end{figure}

Another method for measuring the received flux in individual modes is by interfering the received light with a shaped (i.e. higher order mode) local oscillator on a detector as shown in Figure~\ref{fig:BHD}, preferably a balanced detector. The interference is beneficial because it increases the detection efficiency of the system thanks to the presence of the bright local oscillator. Yang and colleagues first demonstrated this technique using balanced heterodyne detection in 2016 \citep{yang2016far} in which they showed sub-diffraction estimation of the separation of two sources by interfering the received light with a $TEM_{01}$ mode. This method was further demonstrated using 21 modes to perform two-dimensional imaging with resolution twice as good as the diffraction limit \citep{pushkina2021superresolution}. More recently, Gosalia and colleagues analyzed the case of using balanced homodyne detection for the sub-diffraction estimation of two point sources, and applied the theory to both coherent sources (lasers from satellites) and thermal sources \citep{gosalia2023quantum}. Gosalia also analyzed the impact of fixed and fluctuating source centroid misalignment and showed that fixed misalignment is comparatively more detrimental. Around the same time, Xie and colleagues proposed a ``double homodyne" method in which the image light is split and directed to interfere with two HG$_{01}$-shaped local oscillators, one of which is $\pi/2$ out of phase with the other \citep{xie2024far}. Xie and colleagues found that their double homodyne method is better than the single homodyne method for larger source separations. Although single homodyne outperforms double homodyne for smaller separations---for separations $d<<\theta$ the measurement noise of double homodyne detection is approximately twice that of single homodyne---double homodyne still outperforms direct imaging, potentially making double homodyne a more practical super-resolution imaging scheme. However, Yang, Gosalia and Xie all found that homodyne or heterodyne measurements still suffer from a limit analogous to the diffraction limit when imaging weak thermal sources \citep{yang2017fisher,gosalia2023quantum,xie2024far}. Yang, Nair and Tsang attribute this to the vacuum noise inherent in interferometric detection schemes which does not reduce with increased signal, unlike Poisson variance in photon counting \citep{tsang2019resolving}.

Yang and Xie both noted that homodyne/heterodyne detection in multiple higher-order modes would enable full imaging with resolution surpassing the diffraction limit \citep{yang2016far,xie2024far}. However, splitting the incoming signal light to multiple interference measurements does not make efficient use of the available photons. It is interesting to note that this problem was encountered in the context of stellar interferometry \citep{townes2000noise,tsang2011quantum}. A hybrid scheme in which mode sorting is first done by a mode demultiplexer such as an MPLC may overcome this problem though, at present, this has not been investigated.

\subsection{Summary of SPADE-like methods} \label{sec:SPADEsum}
The underlying mechanism of the SPADE imaging method is separation of the incoming modes so that the flux in chosen modes can be measured. The separation of specific modes is what enables the measurement to retain more phase information and thus achieve superior resolution compared to direct imaging.

As described in the preceding sections, numerous technologies for mode sorting have been proposed including MPLCs \citep{boucher2020spatial,tan2023quantum,santamaria2023spatial,rouviere2024ultra,santamaria2024single}, photonic lanterns \citep{ellis2024astrophotonics,birks2015photonic,leon2013photonic,leon2017photonic}, double-clad fibres \citep{de2015double}, image-inversion \citep{tang2016fault,larson2019common,nair2016far,lu2018quantum}, phase plates \citep{tham2017beating}, and holograms \citep{paur2016achieving}, with the sub-diffraction resolution of many of these methods being demonstrated in the laboratory. While these methods all share an underlying relation to SPADE, their specific implementations have advantages and disadvantages for practical application in astronomy, and they have different noise sources that impact their ability to reach the QCRB, most clearly highlighted by the inability of homodyne and heterodyne detection schemes to do so. The use of MPLCs and photonic lanterns seem the most mature and practical since they are available commercial-off-the-shelf, have fixed transformation matrices once calibrated, can be installed at the image plane of optical telescopes with minimal modification, and simply require one single-photon detector per measured mode.

Performing effective parameter estimation (i.e. imaging) of extended or complex sources with any of these SPADE implementations remains a challenge, but could potentially be overcome with neural networks \citep{frank2023passive,pushkina2021superresolution,gozzard2025super,buonaiuto2025machine,sajia2025breaking,kuusela2025optical}, iterative algorithms \citep{matlin2022imaging,desai2025multiple,choi2025super}, or moment-based measurements \citep{sorelli2021moment}.

\subsection{Photon cloning via amplification} \label{sec:nonDem}

\begin{figure}
\centering\includegraphics[width=0.47\textwidth]{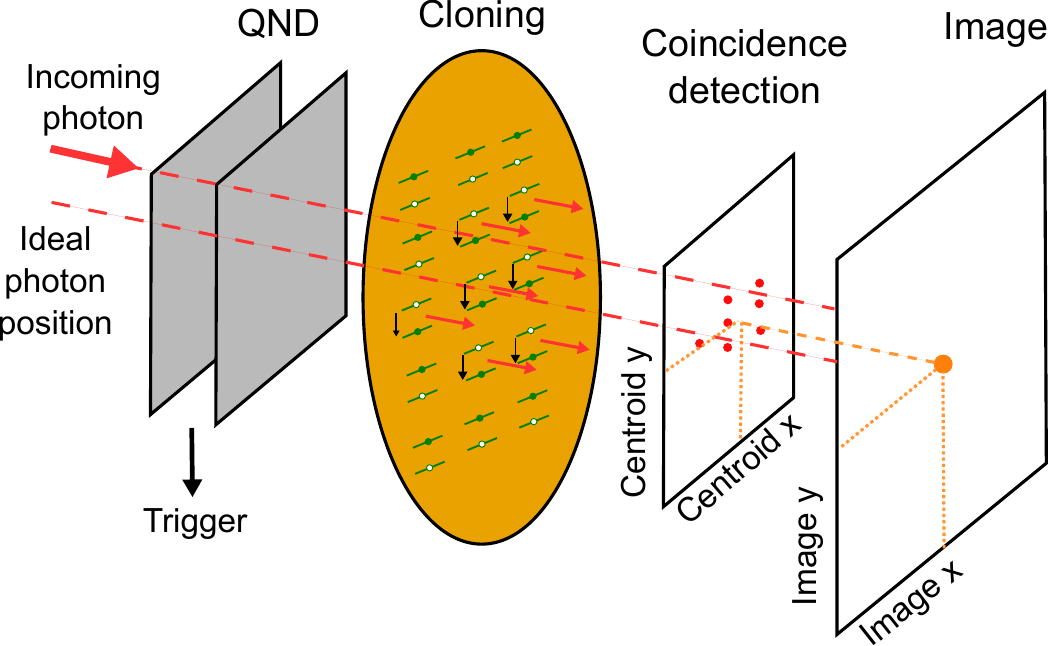}
\caption{Generalised representation of a quantum telescope using quantum non-demolition and cloning. Quantum non-demolition provides a trigger signal to reduce noise due to spontaneous emission from the amplifier. In the amplifier, incoming photons stimulate the production of identical photons, increasing the number of photons received by the detector, thereby increasing the resolution of the imaging system.}
\label{fig:nonDem}
\end{figure}

Kellerer proposed overcoming the diffraction limit by imperfect cloning of incoming photons \citep{kellerer2014beating,kellerer2014quantum,kellerer2016beyond}. Incoming photons stimulate the emission of clones in an optical parametric amplifier. The multiple clones then arrive on a detector that measures the average position of simultaneously arriving photons. The increased number of photons enables improved resolution in a manner analogous to $N00N$ states \citep{boto2000quantum}. A quantum non-demolition measurement is used to reject noise from spontaneous emission from the amplifier.

In a quantum non-demolition (QND) measurement, the photon is detected, but not destroyed \citep{braginsky1995quantum,nogues1999seeing,aspelmeyer2014cavity,xia2016cavity}. An incoming photon interacts with an atom in a highly excited state. The presence of the photon changes the polarization state of the atom, but does not de-excite the atom. Thereby, presence of the photon is detected, but the state of the photon (importantly, its momentum) remains undisturbed. The polarization state of the atom is measured and used to determine the presence or absence of a photon. The received photon is amplified by stimulating the emission of clone photons from excited atoms. \footnote{While perfect cloning is not permitted by quantum mechanics, imperfect cloning is possible and can be used to beat the diffraction limit if the photons share identical momentum states.} The detection of the presence of the original photon triggers a measurement of coincident photons at the detector. In the limit where only cloned photons are measured and stray signals are effectively excluded, the resolution of the telescope would be improved by a factor of $\sqrt{N_C}$ over the diffraction limit, where $N_C$ is the number of clones \citep{kellerer2014beating,kellerer2014quantum}.

At present, such QND measurements and amplification are extremely inefficient. This means that, although this method would surpass the diffraction limit, the sensitivity of the telescope would be extremely poor. Modern QND experiments are performed using photons from intense laser sources due to this inefficiency \citep{kellerer2014beating,kellerer2014quantum}. Furthermore, the method is extremely sensitive to noise introduced by spontaneous emission photons from the amplification stage, which could prevent significant improvements in resolution \citep{stenholm1986theory}. 
It is, in principle, possible to perform noiseless amplification \citep{duan1998probabilistic,ralph2008nondeterministic}, but this technology is not yet mature enough to provide the performance improvement required for effective use of QND and cloning as an astronomical imaging method. Kellerer investigated the possibility of selecting only events where there is an above average ratio of stimulated to spontaneous photons \citep{kellerer2016beyond}. This method is promising, but is also dependent on nascent technologies, and exacerbates the limited efficiency of the imaging system.

Kurke and colleagues performed further analysis and simulation of this imaging method \citep{kurek2016quantum,kurek2018usability}. They concluded that the method is feasible, with the relevant technologies being advanced enough to construct a technology demonstrator in the near future. They also concluded that the method benefits more from increasing the number of clones than from reducing the background noise. Gumpel and Ribak demonstrated a laboratory proof of concept of this imaging method \citep{gumpel2021optical} using solid-state die amplifiers without a QND measurement. In spite of this, they were able to achieve a resolution slightly better than the diffraction limit of their setup.

\subsection{Long baseline quantum interferometers} \label{sec:interferometers}

As noted in Section~\ref{sec:theorySum}, currently the largest optical interferometer is CHARA, with a baseline of 330~m and a maximum resolution of around 300~$\mu$as (1.5~nrad) \citep{pedretti2009imaging}. The next largest interferometer in the optical and NIR wavelengths is the Very Large Telescope Interferometer (VLTI), with a maximum baseline of 130~m \citep{haguenauer2012very}. The GRAVITY instrument on the VLTI can achieve a resolution of around 3~mas (15~nrad) in the NIR \citep{gillessen2010gravity,abuter2017first}\footnote{Astrometry refers to the precise measurement of positions and motions of celestial objects and differs from imaging in that it only requires the optical center-of-mass to be determined. Astrometry can typically be performed to better resolution than imaging with the same instrument. For example, the GRAVITY instrument on the VLTI can perform imaging with a resolution around 1~mas, and astrometry with a resolution of 10-100~$\mu$as.}. A new optical interferometer, the Magdalena Ridge Observatory Interferometer (MROI) is currently under construction. The MROI will have a maximum baseline of 340~m, achieving a resolution of 600~$\mu$as at a wavelength of 1~$\mu$m \citep{buscher2013conceptual,creech2018magdalena}. While this is less than the resolution of CHARA, the MROI's optical design will allow it to image targets five magnitudes (100 times) fainter.

Increasing the baselines of optical interferometers to kilometre-scales and longer (optical-VLBI) has so far proved insurmountable due to \citep{eisenhauer2023advances}:

\begin{enumerate}
  \item \textbf{Low photon flux}. The rate of photons from astronomical sources at visible and NIR wavelengths, even for large 10~m-class telescopes, is on the order of a few photons per second. This makes optical interferometers extremely sensitive to loss in the optics.
  \item \textbf{Phase noise}. An interferometer needs to be phase-stable to at least $\lambda/5$, preferably better than $\lambda/20$. Due to the fact that photons cannot be recorded, at visible and NIR wavelengths, this means the mechanical stability of the interfometer needs to be better than 100~nm. This is extremely challenging.
  \item \textbf{Real-time processing}. The inability to record optical photons for post-processing means that interference measurements must be done in real-time as the signal photons arrive. This exacerbates the issues of low photon flux and sensitivity to phase noise.
  \item \textbf{Wavefront correction and calibration}. Atmospheric turbulence creates phase perturbations on the optical wavefront received by the interferometer, introducing phase noise that reduces the precision of the measurement and this imaging resolution. For optical interferometers comprising multi-meter-class telescopes (such as the VLTI), adaptive optics is required at each telescope to ensure efficient coupling into the optical waveguides. Optical interferometers require \textit{fringe trackers} which use a bright star in the field of view of the interferometer to correct for the effects of atmospheric turbulence in the interference fringe. This limits the astronomical targets available to interferometers to ones where a sufficiently bright guidestar is nearby. Interferometers need to be calibrated prior to observations by observing a source that is beyond the resolution of the interferometer. As interferometers get larger, suitable sources become rare.
\end{enumerate}

Recent advances in quantum and optical metrology technologies have the potential to overcome 
the first three of 
these issues. 
Wavefront correction and calibration will remain an issue, but quantum technologies may offer some advantages over the current state of the art.
For a recent review focusing on quantum-enabled optical interferometers see \citep{huang2026quantum}.

\subsubsection{Optimal imaging with quantum interferometers} \label{sec:phaseStab}

\begin{figure}
\centering\includegraphics[width=0.47\textwidth]{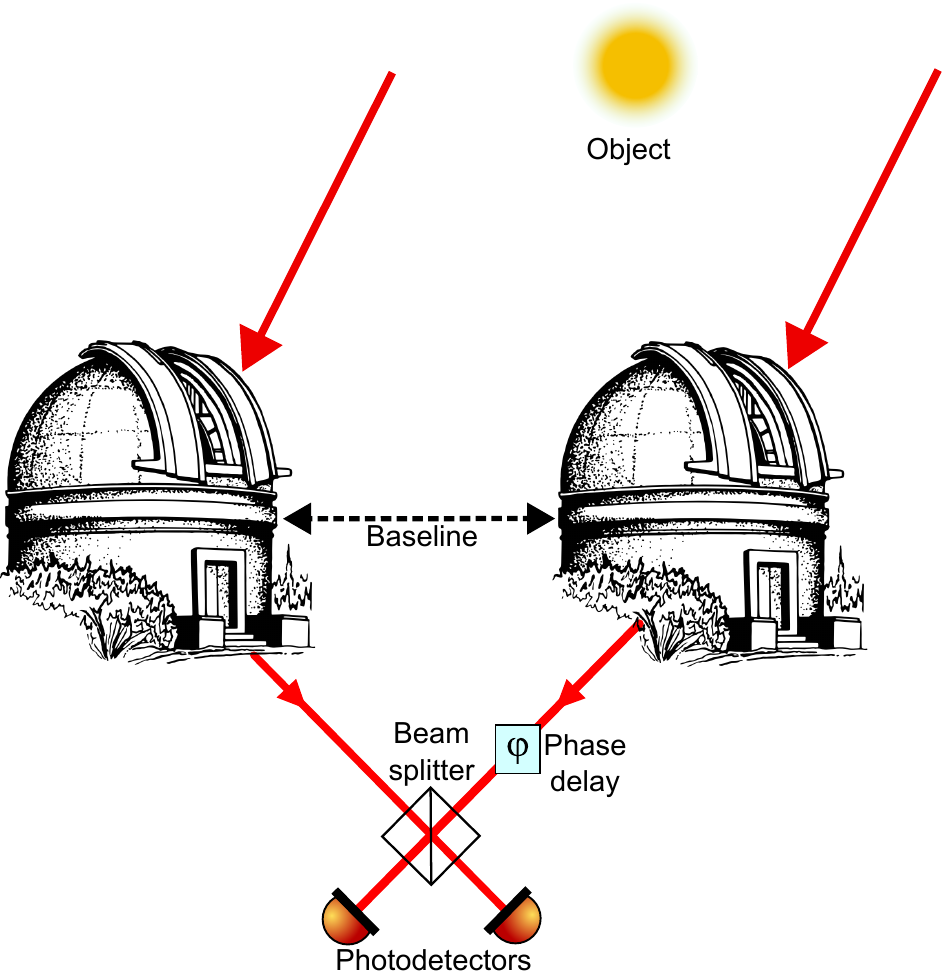}
\caption{Simplified schematic of a two-element astronomical interferometer in which the signal received by the two elements is interfered on a beam splitter. Single-photon detectors at the output of the beam splitter record an interference fringe as the relative phase in the two arms of the interferometer is varied.}
\label{fig:IFO}
\end{figure}

At its simplest, an astronomical interferometer, as shown in Figure~\ref{fig:IFO}, combines the light from two telescopes on a beam splitter. An interference pattern is measured and the visibility function is used to infer the brightness distribution of the source via the van~Cittert–Zernike theorem \citep{mandel1995optical}. 

In 2017, Pearce, Campbell and Kok applied the QFI and QCRB to quantum parameter estimation of distant black bodies \citep{pearce2017optimal}. They showed that their measurement formulation was optimal---equivalent to SPADE for similar numerical apertures---and proposed that the measurement could be achieved with an interferometer, such as shown in Figure~\ref{fig:IFO}, using single photon detectors at the outputs of the beam splitter. Kok and others further showed that such a linear interferometer is always optimal, and that the three-dimensional positions of the sources can be measured \citep{parniak2018beating,lupo2020quantum}.

In the large photon number limit, conventional ``click" single photon detectors that are unable to distinguish between the arrival of one photon or many are able to produce the optimal measurement. However, in the few-photon regime, photon number-resolving detectors are required to perform an optimal measurement. This optimal imaging technique was demonstrated in the laboratory by Howard et~al. \citeyearpar{howard2019optimal} and Zanforlin et~al. \citeyearpar{zanforlin2022optical}.

The ability of quantum interferometers to optimally extract information from the available photons means they offer advantages over classical interferometers in the photon-starved regime typical of astronomical imaging \citep{hu2025superresolution}. As well as offering an efficient way to construct a long-baseline optical interferometer, quantum interferometers also offer super-resolution. By conducting photon counting measurements at the output of the beam splitter, a quantum interferometer achieves better phase sensitivity than a classical interferometer, providing tighter constraints on the measured phase information, and thus better resolution. Additionally, the quantum interferometer is able to take advantage of $N00N$ states. $N00N$ states are superpositions of $N$ photons across two arms of an interferometer \citep{slussarenko2017unconditional,dowling2008quantum}. $N00N$ states produce interference fringes that oscillate faster than the classical interference pattern, enabling the interferometer to achieve phase super-resolution. However, an astronomical interferometer is already operating in a photon-starved regime, and these multi-photon states are going to be rare. It remains to be seen whether exploiting $N00N$ states is realistic for astronomical observations.

Howard et~al. \citeyearpar{howard2019optimal} demonstrated using the interferometer to measure the angular size of a single pseudo-thermal source, and quantified the performance improvement gained in the few-photon regime by using photon number-resolving transition edge sensors \citep{morais2024precisely}. Zanforlin et~al. \citep{zanforlin2022optical} used ``click" single photon detectors, but imaged a pair of pseudo-thermal sources and demonstrated hypothesis testing protocols \citep{huang2021quantum} to determine the presence of a second source and then measure the separation of the sources. Both Howard and Zanforlin's interferometers had short baselines (approximately 5~cm and 5~mm respectively), and both concluded that a major challenge to scaling this quantum interferometer up to baselines long enough to challenge and then displace current optical interferometers such as CHARA and the VLTI was the suppression of phase noise in the arms of the interferometer.

The success of CHARA and the VLTI show that it is already possible for optical interferometers to achieve sufficient phase stability over baselines on the order of 100~m, and the technologies used to do so could be directly applied to quantum interferometers. Alternatively, phase noise suppression techniques for the transfer of stabilised frequency references over long distance fibre optic links \citep{predehl2012920,droste2013optical,schediwy2013high,collier2025phase,zhang2025criteria, collier2026long} could be employed, significantly reducing the complexity of the phase noise suppression systems. These fibre-based phase noise suppression systems are used on radio interferometers including the Atacama Large Millimetre Array (ALMA) \citep{cliche2006applications} and the SKA \citep{schediwy2019mid}, and have already been applied to quantum communications, with various groups demonstrating successful twin-field quantum key distribution (TF-QKD) over links up to 1000~km \citep{clivati2022coherent,pittaluga2021600,bertaina2024phase,liu2023experimental,wang2022twin,pittaluga2025long}. TF-QKD exploits the interference of transmitted quantum states from two sources at a central beam splitter (i.e. a TF-QKD system resembles the quantum interferometer shown in Figure~\ref{fig:IFO} except the quantum states are produced by weak coherent sources rather than astronomical thermal sources) and so is sensitive to phase noise on the links. In these demonstrations, the quantum states co-propagated through the fibre with a bright phase reference signal that was effectively removed by standard telecommunications-grade dense wavelength division multiplexers with minimal impact on the quantum signal. Other studies into the impact of bright classical signal co-propagating in fibre with weak quantum signals find that, although Raman scattering will always introduce noise into the quantum channel, with sufficient attention to the chosen classical and quantum wavelengths, its impacts can be minor \citep{valivarthi2019measurement,dynes2016ultra,schreier2023coexistence,thomas2023designing,thomas2024quantum,ramesh2024hong,linale2024c,stathis2024experimental,nardelli2025phase,jabir2025enabling,nakamura2025sub}.

This gives the authors confidence that a quantum interferometer telescope could be constructed with baselines on the order of hundreds of kilometers with currently available technology. At an observing wavelength of 1550~nm, an interferometer would need a baseline of only 320~km to achieve an imaging resolution of 1~$\mu$as (5~prad), and that is before considering the potential for super-resolution. However, no simulations of what astronomical science is achievable with current technology (i.e. 10~m-class telescopes and wavelengths limited to the telecommunications C-band) have yet been performed, so there remain open questions as to the near-term practicality of this technology for astronomy.

While the fact that signal photons must be used in real time and cannot be stored for later processing, or the processing repeated in the event of an error, is a disadvantage of optical interferometers over radio interferometers, the enormous absolute bandwidths (spanning terahertz) compared with the radio spectrum makes the electronic correlation of signals highly impractical with modern computing resources \citep{bourdarot2021architecture}.

A further challenge for quantum interferometers is efficient scaling up of the array to 2D imaging (requiring at least three apertures) and covering a larger range of the 2D Fourier space to extract more comprehensive spatial information about the source. Pair-by-pair comparison of each baseline within an array of telescopes scales quadratically as

\begin{equation}
    b = \frac{n (n - 1)}{2},
    \label{eq:quadratic}
\end{equation}
where $n$ is the number of elements in the interferometer array and $b$ is the resulting number of baseline pairs. 

For an array like CHARA with six apertures, this is 15 separate baselines, while a radio interferometer array such as the Very Large Array (VLA) with 27 elements requires 351 individual comparisons. Since the astronomical photons cannot be copied, this means the observing time required by the array to produce an image with a particular SNR scales as $b$. More efficient methods are required to make effective use of the available photons and not diminish the sensitivity of the array. Multiport interferometers using quantum Fourier transforms \citep{pryde2003creation,resch2007time} were postulated as efficient solutions to this problem. Recently, Zhang~et~al.~\citeyearpar{zhang2026quantum} showed that a properly arranged interferometer with multi-aperture interference can achieve quantum-optimal sub-diffraction measurement of image moments up to $2N-2$ where $N$ is the number of apertures. Furthermore, they show that their arrangement is equivalent to a synthesised aperture version of SPADE imaging. As a result, they dub this arrangement ``array-SPADE''.

Some consideration has been given to hybrid interferometers that combine aspects of SPADE with interferometry \citep{sajjad2024quantum} where the de-multiplexed spatial modes from separate apertures are directed to linear interferometers, extracting additional information from the measurement.

\subsubsection{Interferometers exploiting entangled or ancilla photons} \label{sec:gottesman}

Quantum repeaters and entanglement sharing offer the possibility of significant advantages for astronomical interferometers over designs that rely on propagation of the astronomical photons to the beam splitter via many kilometers of optical fibre. By interfering astronomical photons arriving at the telescopes with photons transmitted to the telescopes from a central entangled photon source, the loss of the fibre links can be absorbed by the entangled source, rather than causing loss of signal photons. The shared states provide a phase reference that, following interference with the astronomical photons arriving at the telescopes, is equivalent to interfering the astronomical photons together \citep{bojer2022quantitative}. In principle, the scheme can be extended to very long baselines by exploiting entanglement sharing between nodes in the distribution network.

This scheme was first proposed by Gottesman, Jennewein, and Croke in 2011 \citep{gottesman2012longer}. Others have expanded on the technique \citep{shen2025photon}, including showing that using a quantum Fourier transform provides improved signal-to-noise ratio compared to classical processing of the interference fringe, an information compression method that improves the efficiency of the imaging by requiring interactions with fewer entangled photon pairs \citep{marchese2023large,khabiboulline2019optical,khabiboulline2019quantum}, and combining SPADE with interferometry by siting a mode demultiplexer at each aperture and performing mode-wise interference using distributed entanglement \citep{padilla2026superresolution}.

Brown et~al. demonstrated this method in the laboratory where they recovered the spatial autocorrelation of quasi-thermal double-slit sources in a single spectral-temporal mode where the single photon was produced by heralded parametric down conversion \citep{brown2023interferometric}, and have proposed performing an on-sky demonstration of the technique using the Sun \citep{brown2024demonstrating}.
More recent demonstrations include using entangled quantum memories to perform imaging over a 1.55~km effective baseline \citep{stas2025entanglement}, and over a 20~km effective baseline while also using the memories to compensate a 1.5~km imbalance in the interferometer arms \citep{wang2025memory}.

However, to create practical interferometer telescopes, the entangled photon source would need to generate on the order of 100~million entangled states per second over a bandwidth on the order of 100~nm in the NIR. (Others have calculated that only a few hundred photons would be needed \citep{leng2024piecemeal}, but these analyses are possibly underestimating the needs of practical astronomical observations.) This is orders of magnitude more than can be achieved with current entangled sources \citep{santra2020entanglement}, meaning that quantum interferometers exploiting entanglement sharing for astronomy are not possible in the near- or medium-term. In addition, in spite of the apparent advantages conferred by transmitting a phase reference out to the receiving telescopes to mitigate the loss of astronomical photons through long fibre links, it turns out that heterodyne interferometry is fundamentally inferior to direct interferometry \citep{tsang2011quantum,michael2023ab,huang2024limited,purvis2024practical,muratore2025superresolution}.

\begin{figure}
\centering\includegraphics[width=0.47\textwidth]{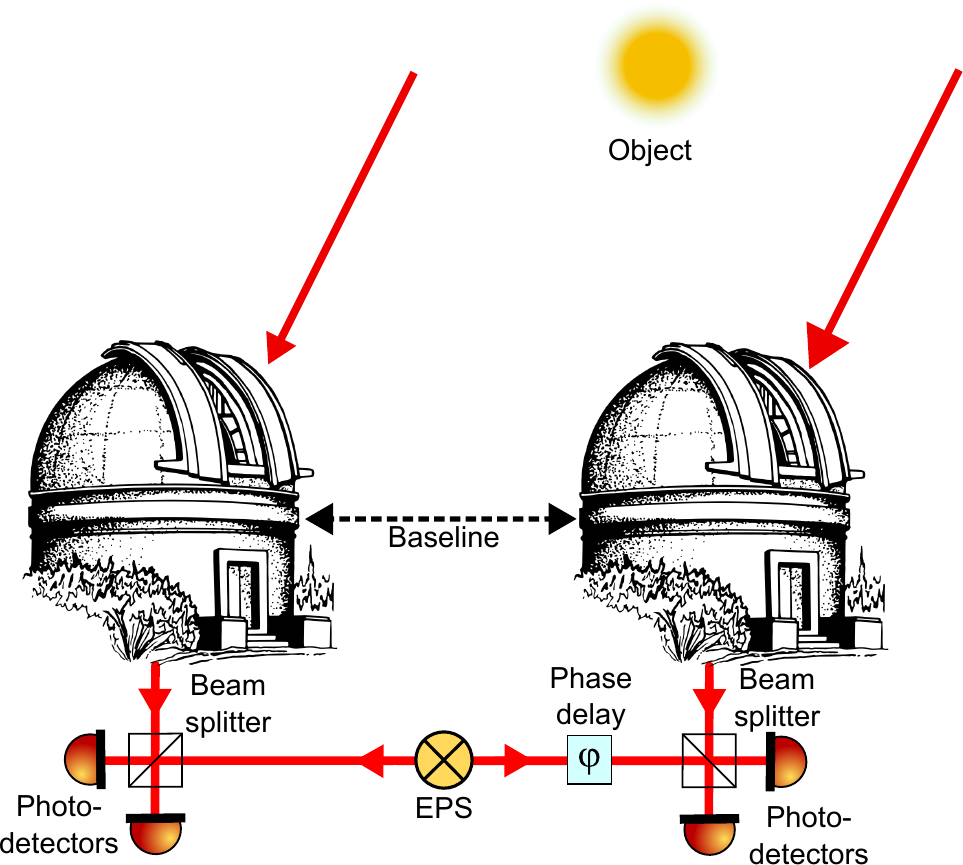}
\caption{Simplified schematic of a two-element astronomical interferometer in which the signal received by the two elements is interfered locally against photons distributed from an entangled photon source (EPS).}
\label{fig:EPS_IFO}
\end{figure}

A similar process proposed by Marchese and Kok \citeyearpar{marchese2023large} using distributed ancilla photons instead of entangled states may overcome this limitation thanks to the greater brightness and bandwidth of suitable single photon sources. Compared with interferometers using entanglement distribution, interferometers using ancilla photons have the advantage of reduced vulnerability to fibre loss and are better suited to the repeaterless quantum networks that can be achieved in the near term. While Modak and Kok \citeyearpar{modak2025large} later showed that this method is very resilient to imperfect indistinguishability of the ancilla photons, the sensitivity of the telescope quickly deteriorates with higher ancilla numbers due to the low photon occupancy rate (faintness) of the starlight.

\subsubsection{Quantum hard drives} \label{sec:qDrives}

An advantage of radio VLBI is that electronics are sufficiently fast to record the amplitude and phase of the received signal, allowing interferometetry to be performed offline. Many VLBI results are obtained by physically transporting hard drives from the receiving elements to a facility where the signals are synchronised and correlated. This process could become feasible for optical interferometers if quantum memories become sufficiently robust to reliably store many thousands of photon states for periods of hours while the memory is physically transported by road or aeroplane.

Bland-Hawthorn and colleagues proposed using quantum memories to non-destructively preserve the photonic quantum states at each interferometer element. The quantum memories can then be brought together and the outputs interfered. They refer to this as a quantum hard drive.

Compared with using quantum repeaters to transmit states between telescopes for interference, Bland-Hawthorn et~al. point out that quantum repeaters are an even less mature technology than quantum memories, requiring significant improvements in single photon sources, entangled photon pair sources, Bell state measurement circuits, \textit{and} quantum memories to feasibly support a large-scale quantum interferometer \citep{bland2021quantum}. 

Quantum memories using optically addressable nuclear spin states in Eu$^{3+}$-doped crystals have been demonstrated to have coherence lifetimes in excess of six hours \citep{zhong2015optically}. This coherence time means that transporting the quantum memory at speeds greater than 10~km/h would achieve lower losses than propagation of the photons through fibre \citep{bland2021quantum}. However, the transportability of these memories has not yet been demonstrated. While the transportability of such quantum memories is likely to be demonstrated in the near future, the bandwidth of the memory is small by astronomy standards, only around 50~MHz, severely limiting the sensitivity of the imaging system. In addition, the storage density of quantum memories will need to be significantly improved in order for a sufficient number of states to be transported. Currently, memories such as the one demonstrated in \citep{zhong2015optically} can stored a few hundred to a few thousand quantum states. Hundreds of millions would need to be stored for a practical astronomical imaging system.

Because of these challenges, quantum interferometers employing transportable quantum hard drives are unlikely to appear in the near future. However, current quantum memory technology may be useful when applied as quantum state delay lines for the type of quantum interferometer described in Section~\ref{sec:phaseStab} \citep{wang2025memory}. Looking further ahead, if high-capacity quantum memories can be deployed on satellites, and down linked via quantum communications, quantum memories could enable space VLBI, such as used in the RadioAstron program \citep{kardashev2014orbit}, with baselines beyond what is possible with Earth-based telescopes.

\section{Astronomical science goals} \label{sec:astroGoals}

The resolution, sensitivity, and observing wavelengths of future telescopes are defined by their scientific goals. These goals are required to build a science case for the financial investment in the construction of the telescope. It is thus necessary for the quantum imaging community to be familiar with the potential science cases for quantum telescopes in order to both direct research efforts towards practical large-scale quantum telescopes, and to support proposals for future instruments. 

Over the past 30-40 years, the advances in astronomical instrumentation to make high resolution observations of objects that have small angular extents and brightnesses have been best benchmarked by the proliferation of the study of exoplanets. It is now well known that many of the stars in our local neighbourhood are surrounded by planetary systems, with over 6400 exoplanets now identified at the time of writing in 2026\footnote{\href{}{https://exoplanet.eu/catalog/}}. Of these, over one thousand systems are comprised of multiple planets orbiting the same host star. Many of these discoveries have relied on indirect observations, deducing the presence of exoplanets through small variations in the positions of the central star over time \citep{Hatzes2016}, or through the periodic brightening and dimming of the host star as the exoplanet occults the stellar surface \citep{Deeg2018}. Such studies are subject to significant uncertainties and degeneracies that limit our ability to infer exoplanet mass and composition with a high degree of accuracy. These challenges can be overcome using direct imaging \footnote{Note that, in astronomy, ``direct imaging" means to directly detect photons from the target object, rather than inferring its existence through other observations. In the section, we will refer to ``direct imaging" in full to differentiate it from the quantum imaging sense (referred to in the previous sections as ``DI") referring to classical imaging.}. Direct imaging has demanded the development of cutting-edge techniques, like extreme adaptive optics and the construction of larger and larger telescopes to enable observations with $\sim\mathrm{mas}$ resolution required to obtain photons that directly convey information about the planetary system \citep{Claudi2024}. The introduction of quantum imaging would bring about a new revolution in exoplanet imaging. In particular, at a resolution of 1~$\mu$as it would be possible to resolve the angular size of planets in the most nearby star systems, for example the planets that are known to orbit the closest stars to Earth \citep{Wagner2021}. A quantum telescope can also probe the structure of protoplanetary disks to reveal how planets -- including our solar system -- form from stellar nebulae \citep{Andrews2020}. Nevertheless, imaging of exoplanets systems also presents an additional but welcome engineering challenge to the development of quantum imaging technologies as it relies on high-contrast imaging (HCI): there is a typical factor of $\sim10^8$ difference in brightness between the planet and the star (see \cite{Zurlo2024} for a review of the limitations and opportunities of modern HCI techniques).\\
At lower contrasts, there are still abundant targets to demonstrate the incredible value of quantum imaging to astronomical science. Understanding the processes that occur during star formation are vital to understand how our Universe evolves \citep{Mckee2007}. However, our ability to observe the collapse of gas clouds and the initiation of stellar fusion in the star formation process is limited by the spatial resolution of our telescopes at infrared wavelengths. Alternative methods that utilise spectroscopy \citep[e.g.][]{Movsessian2024} can be used to observe these processes through the outpouring of material to form $\sim\mathrm{pc}$ scale shock-excited nebulae that are generated by newborn stars \citep{Herbig1951, Haro1952, Haro1953}. High angular resolution imaging would provide a key opportunity to peer closer into the center of collapsing interstellar gas and dust clouds and catch star formation in the act.\\
Another key science goal is to spatially resolve the surfaces of nearby stars, allowing us to significantly advance our knowledge of stellar physics. Directly imaging the surfaces of stars beyond our Sun can reveal the role of angular momentum in stellar evolution, the role of magnetic fields and how stellar winds are launched from massive stars, as well as how the interactions between stars with binary companions influences their evolution and properties, improving on results obtained via Doppler imaging \citep{Luger2021}. Optical VLBI has allowed for the surface of rapidly-rotating star Altair to be imaged with a resolution of $<1\,\mathrm{mas}$, revealing a stellar photosphere that was highly distorted, with significant gravity darkening that could only be explained by either differential rotation or an alternative gravity darkening law \citep{Monnier2007}. Observing stars like Altair with the higher spatial resolution could reveal further details of the physics of stellar photospheres and allow us new ways to test our theories of stellar evolution. The ability to observe stellar surfaces at high angular resolution would also enable the imaging of starspots, a phenomenon that has been proposed by  \cite{Alexeeva2021} as an explanation for the recent dimming of red supergiant star Betelgeuse. Observations of starspots reveal the magnetic properties of stars \citep{Roettenbacher2016}, however existing observations of starspots are limited to indirect spectroscopic observations and Doppler imaging. These techniques have limitations, notably the misinterpretation of noise as starspots close to the stellar poles \citep{Brus1998}, which can be overcome with high resolution quantum imaging.\\
Quantum imaging can reveal the multiplicity of stars in the local Universe. $\sim70\%$ of all O-type stars are expected to interact with a companion at some point \citep{Sana2012, Moe2017}. The most accurate way of determining if a star is part of a close binary system -- where the photospheres of the stars interact with one another in such a way that it has a strong influence on the evolution of the star, its nucleosynthesis processes and magnetic fields -- is through direct imaging of the stellar surface. With the resolution of a quantum telescope, it would be possible to directly probe the intrinsic binary fraction of stars in our Galaxy. There are numerous known contact binaries that can be readily targeted using a quantum telescope sensitive in the infra-red \citep{Gettel2006}.\\
The innermost kiloparsec of our Galaxy offers a rich selection of astrophysical phenomena \citep{Morris96gcreview, Genzel2010rev} whose study can be substantially enhanced by the improvements in resolution offered by quantum imaging. The central supermassive black hole (SMBH) Sgr A* was initially identified as a persistent radio source of unknown origin \citep{Balick1974disco}. More recently, radio VLBI observations with the Event Horizon Telescope \citep[EHT, ][]{Doelman2009EHT} have enabled the direct imaging of structures close to the black hole event horizon: observations at a wavelength of $1.3\,\mathrm{mm}$ reveal a bright ring of material orbiting the black hole with an angular diameter of $51.8 \pm 2.3\,\mu$as~\citep{EHTsgrA}. Improvements in the resolving power of telescopes through quantum imaging could enable more precise observations of the shape of the black hole shadow and any substructure in the ring of emission resolved by EHT to corroborate and refine the findings of the EHT. In particular, a quantum imaging system could allow higher precision observations of the accretion disk structure to uncover more information about how Sgr A* accretes material, produces flares, and how the ring structures observed by EHT vary with time. Such a telescope would complement the capabilities of the EHT at $\sim \mathrm{mm}$ wavelengths. Observations to constrain the morphology of the black hole shadow in particular would allow for stringent tests of GR, and allow us to search for evidence of any modified theory of gravity in the strong field limit \citep{Uniyal2025BHshadow}.\\
However, well before these advances in imaging that allowed us to peer directly at Sgr~A*’s accretion disk, it was the determination of the orbital ephemeris of the stars that occupy Keplerian orbits in the innermost parsec of the Galaxy that revealed Sgr~A*’s identity as a SMBH with $M_\mathrm{SMBH} \sim 4\times10^6\,\mathrm{M_\odot}$. The `S-stars' of the nuclear star cluster were observed over decades-long campaigns \citep{Schoedel2002GC, Ghez2005, Gillessen2009, Gravity2018red} as they orbit a point coincident with the Sgr A* radio source revealed the presence of a $\sim 4\times10^{6}\,\mathrm{M_\odot}$ supermassive black hole, and it was this work using the S-stars that yielded the award of the 2020 Nobel Prize for Physics to Reinhard Genzel and Andrea Ghez (along with Roger Penrose for his theoretical work on black holes)\footnote{https://www.nobelprize.org/prizes/physics/2020/summary/}. \\
How nuclear star clusters form and evolve is not well understood \citep{Neumayer}, and the Galactic Center provides an excellent opportunity to explore their structure, evolution and origins in our own cosmic back yard. High resolution imaging could not only improve our orbital ephemeris and yield a more precise estimate of the mass of Sgr~A*, but can also shed further light on the origins of the nuclear star cluster and any ongoing star formation in that region. Due to high levels of dust extinction at optical wavelengths, the stellar population of the Galactic Center is naturally best studied at $\sim 1-2$~$\mu m$ \citep{Becklin68}. \\
As well as enabling observations of the accretion disk that forms around our galaxy’s central SMBH, it is possible that quantum imaging could allow us to probe accretion on smaller size scales. High angular resolution observations could enable us to resolve the accretion streams that form in eruptive variables, low-mass x-ray binaries, and novae - optical emission from accretion streams may be visible for several days before these objects emit the x-rays they are usually detected with~\citep{Russell2019}. At the highest angular resolutions, it may become possible to probe the behavior of accretion disks close to the surfaces of neutron stars, to study the interplay between the physics of the accretion disk and the influence of space-time curvature close to the compact object where infrared emission arises from cyclo-synchrotron radiation from extended hot gas in the accretion stream \citep[e.g.][]{Vincentelli2023}. \\
Finally, there is also abundant opportunity to extend the scope of the advancements that can be made through quantum imaging to extragalactic sources. Most notably, this could involve imaging the central parsec of nearby galaxies to provide definitive evidence for the existence of close, sub-parsec SMBH bianries in the nearby Universe \citep{SMBHbinarieschapter}, and to resolve the inner parsec from which powerful jets are launched by SMBHs like M87*~\citep{Prieto2016}.

\subsection{Useful targets for proof-of-principle tests}
As quantum imaging technologies for astronomy advance beyond laboratory demonstrations, on-sky tests using actual astronomical targets will be necessary to prove the technology. Technologies suitably advanced to be tested at scales on the order of 10~m to 100~m can be tested using existing telescopes and interferometers. In these cases, targets at or just beyond the resolution of the original instrument that have previously been observed will provide a range of suitable candidates. However, for small-scale or proof-of-principle testing of technologies, there is a limited range of objects suitable for study.

\begin{itemize}
  \item \textbf{Solar system objects}. The Sun may be needed to test low-TRL technologies that require exceptionally bright sources, such as interferometers using EPS \citep{brown2024demonstrating}. Jupiter and Saturn present very bright sources suitable for testing the resolution and measurement of their systems with small-aperture telescopes. Jupiter's Galilean moons and Saturn's rings (and moons) also present useful targets for discriminating the presence, size, and angular separation of smaller secondary objects, or for measuring the optical moments of the system.
  \item \textbf{Extra-solar objects}. Several star systems present useful targets. Betelgeuse (Alpha Orionis) is the star with the largest angular diameter from Earth (approx. 50~mas), and is the brightest star in the night sky in the astronomical H-band (encompassing 1550~nm, the H-band is centered on 1630~nm and has a FWHM of 307~nm) with a magnitude of $-$3.73 \citep{ducati2002catalogue}. This makes Betelgeuse a good target for measurements of angular diameter and limb-darkening. Sirius (Alpha Canis Majoris) is the brightest star in the night sky in the visible band. Its large angular diameter from Earth (5.9~mas) also makes it a suitable target for measurements of angular diameter and limb-darkening. The presence of a secondary white dwarf companion (Sirius B) 9,124 times fainter than the primary provides a useful test of protocols looking for faint secondary companion objects such as exoplanets. In the H-band, Sirius is approximately 1/9$^{th}$ the brightness of Betelgeuse with a magnitude of $-$1.33 \citep{ducati2002catalogue}. Beteleguse and Sirius are visible from both the northern and southern hemispheres, making them useful targets for comparison by different research groups. Alpha Centauri is a triple system with Alpha Centauri A and B being separated by approximately 5.5~as. Alpha Centauri A (H-band magnitude $-$1.886 \citep{cutri20032mass}) is approximately 3.6 times brighter than Alpha Centauri B (H-band magnitude $-$0.49 \citep{ducati2002catalogue}). Alpha Centauri has a combined magnitude in the H-band of $-$2.15, so is brighter than Sirius in the H-band by a factor of approximately 2.1. This makes Alpha Centauri a good target for tests of protocols discriminating the presence of a secondary companion of comparable brightness. However, Alpha Centauri is only visible from the southern hemisphere.
\end{itemize}

\section{Challenges and future technologies} \label{sec:challenges}

While the technologies and experimental demonstrations discussed in Section~\ref{sec:q4astro} provide a strong indication that various quantum imaging methods have application in astronomy, there are key challenges that future development will need to address in order for these technologies to find widespread practical use in astronomical imaging.

\subsection{Imaging extended sources} \label{sec:extendsource}
While the estimation of the separation and brightness of two sources has some uses in astronomy---possibly most significantly, exoplanet detection---its practical applications are limited. Quantum imaging methods suitable for arbitrary sources will find wider utility in astronomical observations. However, as noted in Section~\ref{sec:spade}, the complexity of full 2D imaging of an arbitrary field with SPADE is prohibitive, and further work into neural networks \citep{frank2023passive,pushkina2021superresolution,gozzard2025super,buonaiuto2025machine,sajia2025breaking,kuusela2025optical}, iterative algorithms \citep{matlin2022imaging,desai2025multiple}, and moment-based measurements \citep{sorelli2021moment} will be required to create a practical system for imaging arbitrary or complex targets.

\subsection{Extending interferometers to 2D} \label{sec:2dinterfero}

Similarly, interferometers with many apertures will be needed to image arbitrary sources. However, as noted, this is not currently achievable without a quadratic increase in the required integration time to achieve required SNRs in each baseline. Theoretical and experimental investigation of multiport interferometers using quantum Fourier transforms may provide a solution to this problem \citep{pryde2003creation,resch2007time,zhang2026quantum}.

Even without these more advanced methods, interferometers comprising at least three apertures are required to determine the \emph{closure phase} of the array. Closure phase enables self-calibration of the array against phase errors, and is resilient to the effects of atmospheric turbulence \citep{eisenhauer2023advances}.

\subsection{Adaptive optics} \label{sec:AO}

\subsubsection{Individual apertures}
Adaptive optics (AO)---a mechanism by which the effects of atmospheric turbulence (called the atmospheric \emph{seeing}) on the wavefront reaching the image plane is corrected---are required to achieve resolution closer to the diffraction limit of the telescope for telescopes with apertures larger than about 2~m. There is little reason to suggest that quantum imaging methods, other than HBT intensity interferometry, will be less sensitive to atmospheric seeing than direct imaging. While HBT intensity interferometers are largely robust against atmospheric seeing, AO will be required for any large-aperture telescopes to achieve good sensitivity.

There is no obvious reason that quantum imaging methods such as SPADE could not make use of existing AO technology with little impact on the quantum imaging if care is taken to exclude the light of the guidestar (either natural or artificial) from the optical path of the quantum system.

Alternatively, photonic lanterns have been successfully demonstrated for simultaneous imaging and image-plane wavefront sensing for AO, with key advantages over more conventional pupil-plane wavefront sensors \citep{norris2020all,norris2024photonic,eikenberry2023photonic,kim2025sky}. The similarity between photonic lanterns and MPLCs strongly suggests that MPLCs could also be used for this method, making the MPLC (or photonic lantern) its own AO wavefront sensor as well as imaging system.

\subsubsection{Interferometers}
In addition to AO, interferometer telescopes also need a way to detect and compensate for time-varying optical delays due to atmospheric turbulence. This is called a \emph{fringe tracker} and has successfully been demonstrated on current optical interferometers using a technique called dual-beam interferometry. In dual beam interferometry, a bright guidestar is used as a reference to stabilise the interference fringe and allow imaging of a fainter object in the same field of view. The technique enabled optical interferometers such as the VLTI to extend coherent signal integration times by three orders of magnitude from around 50~ms to more than 50~s, allowing much fainter magnitudes to be reached \citep{eisenhauer2023advances}.

Quantum interferometers should be able to make use of the dual-beam interferometry technique, but care needs to be taken to minimise the noise introduced to the quantum channel from the bright guidestar.


\subsection{Single photon detectors} \label{sec:SPDs}
Single photon detectors with high efficiency and low dark counts at wavelengths of 2~$\mu$m and longer are required to open up the astronomical K-, L- , M-, N-, and Q-bands to quantum imaging techniques. Superconducting nanowire single photon detectors (SNSPDs) with good performance have been demonstrated at wavelengths up to 10~$\mu$m \citep{wollman2021recent,dello2022advances}. However, SNSPDs are polarization sensitive, so polarization beam splitting of the astronomical photons will be required, which may introduce significant noise into the measurement.

\subsubsection{Counting photons}
As shown in Howard et al. \citeyearpar{howard2019optimal} and discussed in Section~\ref{sec:interferometers}, photon number-resolving single photon detectors will be required to exploit the full super-resolution capabilities of interferometers, though this will only be of benefit in extremely photon-starved regimes. SNSPDs can be manufactured to be photon number-resolving \citep{stasi2023fast}, but another technology that may have advantages over SNSPDs is transition edge sensors (TES) which are a type of calorimeter able to discern the number of detected photons by the total energy deposited in the superconducting material \citep{morais2024precisely}. The results in \citep{howard2019optimal} suggest that photon counting only works for near-monochromatic light. This may mean that multiple detectors are required to cover a single astronomical observing band. It remains to be seen whether the relative rarity of multi-photon coincidence events makes photon number-resolving detectors of practical value in astronomical imaging.

\subsection{Interferometers}

Many of the difficulties in scaling up classical optical interferometers to larger baselines (i.e. $>$~1~km) will apply to quantum interferometers. Concepts for astronomical optical interferometers larger than 1~km in baseline have already been proposed \citep{jovanovic20232023} and these roadmaps detail the challenges and technologies needed to build such instruments. CHARA is currently commissioning a mobile fibre-linked telescope into the array \citep{scott2024cmap,koehler2024integrating,magri2025first} with notable success. Quantum astronomical interferometers will benefit from the advances in classical interferometers and it is likely that quantum interferometer telescopes will initially be demonstrated as instruments retrofitted to classical interferometers.

In addition to the challenges faced by classical interferometers, quantum interferometers face additional challenges for which technologies or solutions must be developed.

\subsubsection{Low loss fibres}
Optical fibre links provide the only practical solution to connecting optical interferometers over distances in excess of 10~km, however, the transmission bandwidth of optical fibre is extremely narrow by astronomical imaging standards. The astronomical H-band, centered on 1630~nm, spans 307~nm, while the telecommunications C-band spans only 35~nm from 1530~nm to 1565~nm. The low-loss wavelength region of optical fibre spans 1260~nm to 1625~nm, covering a significant fraction of the astronomical J- and H-bands. While this total bandwidth is now comparable to the width of astronomical observing bands in the NIR, it represents only a single band, limiting the range of astronomical science that can be done with interferometers connected via modern fibres. In addition, the dispersion of these wavelengths through the fibre represents a significant technical issue that must be addressed. The effectiveness of phase stabilization of the interferometer over wide bandwidths will also need to be investigated \citep{bertaina2024phase}. It should be possible to counteract the dispersion of the wide-band astronomical signals within the fibres with dispersion-compensating elements prior to interference \citep{Litchinitser2007,Treacy1969optical}.

Building practical large-scale interferometers at longer wavelengths will require optical fibre with low losses in the mid infrared, above 2~$\mu$m, in order to open up the most scientifically significant observing wavelengths. A variety of fibre waveguide technologies already exist with this capability, most notably hollow waveguides \citep{jackson2021mid}. Hollow core fibres have achieved losses of 0.091~dB/km \citep{petrovich2025broadband}, about half that of conventional single mode fibre, with significant potential for improvement. Hollow core fibres are also much more broadband than single mode fibres, exhibit significantly lower dispersion of nonlinear effects because the light propagates through air instead of solid glass, and have the potential to support wavelengths as long as 5~$\mu$m \citep{hecht2026beyond}. Another exciting recent development is ZBLAN fibre, which has a theoretical loss coefficient of 0.006~dB/km \citep{moser2026flawless,aratspace2026,tucker2020production}, more than an order of magnitude better than conventional fibre.

At present, these fibres are relatively expensive to manufacture compared with standard telecommunications SMF-28, however, the length of fibre needed to construct a 10~km- to 100~km-scale interferometer would likely be feasible within the construction budget of such an observatory.

\subsubsection{Single photon sources}
As noted in Section~\ref{sec:gottesman}, using entangled photon sources to perform entanglement swapping between distant apertures would improve the sensitivity of the instrument by avoiding the loss suffered by propagating the astronomical photons through long lengths of optical fibre. However, to create practical interferometer telescopes, the entangled photon source would need to generate on the order of 100~million entangled states per second over a bandwidth on the order of 100~nm, a requirement that is currently out of reach of EPS technology \citep{santra2020entanglement}. Much effort in the development of suitable entangled sources will be needed if astronomy is to benefit from this technology.

Also as noted in Section~\ref{sec:gottesman}, single photon sources suitable for generating ancilla photons will likely meet these requirements in the nearer term \citep{somaschi2016near}, with additional advantages over EPS-based interferometers in terms of resilience to fibre loss and compatibility with near-term quantum networks.

\subsubsection{Quantum memories}
Optical interferometers require delay lines to compensate for the geometric delay of the arrival of photons at different apertures. Current interferometers including CHARA, the VLTI and the MROI use retroreflectors mounted to carts to continuously adjust the delay in real time \citep{eisenhauer2023advances}. This is unlikely to be practicable for interferometers exceeding 1~km baselines. 

For interferometers spanning on the order of 10~km to 100~km, near-term quantum memory technology may prove suitable for use as delay lines. An interferometer spanning 320~km would require approximately 1~ms of delay to compensate for the geometric path length difference. Current quantum memory technology can store and release quantum states for this period with an accuracy better than 70\% \citep{lei2023quantum}. If the storage time of the quantum memory can be continuously adjusted between 0~ms and 1~ms with sufficient precision, using quantum memories as optical delay elements would avoid the need for hundreds of kilometers of mechanical or fibre delay lines.

The storage volume requirement of the quantum memories would also be relatively manageable. Astronomical photon rates for a typical observation will be on the order of 10,000 photons per second, meaning the quantum memory would need to store only on the order of 1,000 quantum states. This is within the reach of current quantum memory technology \citep{lei2023quantum}.

\subsubsection{Calibration and image reconstruction}
Interferometers require phase calibration routines to be performed in order to correctly synchronise the phase delays from each aperture \citep{grainge2017square}. This is typically performed by pointing the array at a calibration source: a bright source that is smaller than the resolution of the interferometer. However, at 10~$\mu$as  and 1~$\mu$as resolutions, especially for such sparse arrays as is typical of optical interferometers, sufficiently bright but compact sources for calibration become impossible to find, necessitating other methods of calibration. Extremely long baseline radio interferometers such as the EHT already face this problem, and employ iterative calibration procedures to calibrate the array \citep{akiyama2019firstIII}.

Image reconstruction is another challenge. Images produced from sparse arrays with low coverage of the u-v plane will feature a large number of artifacts due to the gaps in Fourier frequency coverage. These artifacts can be easily removed for simple sources such as a small number of compact objects \citep{hogbom1974aperture}, but extended or complex sources require a forward modeling approach that considers all possible astronomical images as hypotheses and then chooses the best set of images based on the best fit to the observables and prior information. Again, 
suitable algorithms and processes have already been developed by the radio and optical interferometry communities \citep{baron2016image,Baron2020inference}.




\section{Conclusion} \label{sec:conclusion}

Recent advances in both quantum information theory as applied to imaging, and quantum metrology technologies, hold great promise for application to astronomical imaging, where optimally-designed measurements can still extract information beyond that available to classical direct imaging methods and achieve super-resolution, in spite of the fact that the optical sources are thermal in nature and cannot be influenced by the observer. Such \emph{quantum telescopes} would have resolutions orders of magnitude beyond current visible and infrared wavelength observatories, and even surpass the angular resolution of radio VLBI networks such as the EHT. This capability would usher in a new era of astronomical discovery, potentially including detailed measurements of black hole accretion disks, the regions around neutron stars, and direct observation of exoplanets.

SPADE imaging, enabled by MPLC technology, holds great promise for improving the resolution of individual telescopes (single apertures) especially if MPLCs with very low crosstalk can be produced, or if noise mitigation and error correction protocols can be implemented \citep{sakuldee2024suppression}. SPADE devices such as MPLCs can be installed on existing and future telescopes with little modification, simply requiring the MPLC to be placed where an astronomical camera normally would be. The greatest current challenge to SPADE imaging is the development of protocols and algorithms to accurately recover the distribution of complex or arbitrary sources where models of the sources are limited.

An improvement in resolution by a factor of 100$\times$ the diffraction limit is reasonably possible with MPLC-based SPADE. On a 40~m-class telescopes such as the ELT, this would enable an angular resolution of 50~$\mu$as, opening up new avenues in the study of stellar surfaces, stellar winds, circumstellar disks, and the environment surrounding the SMBH at the center of our galaxy.

Possibly most exciting is the prospect for application of quantum metrology techniques to astronomical interferometers. To achieve an angular resolution of 1~$\mu$as (4.85$\times 10^{-12}$~rad) an interferometer observing in the 1-2~$\mu$m range would need to span a baseline of 200-400~km. This is possible with current single photon detectors and phase stabilization technologies. Even if it proves impractical to perform super-resolution imaging with large-scale interferometers, the technologies discussed in this review would still enable the construction of these long-baseline interferometers for classical interferometric imaging at 1~$\mu$as resolutions.

Quantum imaging offers feasible methods to achieve 1-10~$\mu$as resolution astronomical imaging at visible and infrared wavelengths without resorting to space-based interferometers or larger space telescopes. However, the quantum technologies presented in this Review will also benefit larger space telescopes if they are realised.

The clear scientific potential of quantum telescopes makes them a promising option for the instrumentation development priorities of the international astronomical community following the current generation of 40~m-scale apertures and 100~m-scale interferometers. To that end, the authors propose the focus be on the development of a 300~km to 400~km baseline fibre-linked interferometer, observing at 1.5~$\mu$m to 2~$\mu$m wavelengths, incorporating fibre-based phase stabilization and delay lines facilitated by quantum memories. Hollow core or ZBLAN optical fibres will greatly improve the efficiency and performance of such an array.

\section*{Acknowledgments}
This work was supported by the Australian Research Council Centre of Excellence for Engineered Quantum Systems (CE17010009). This material is based upon work supported by the Air Force Office of Scientific Research under award number FA2386-23-1-4081. D.R.G. is supported by a DECRA Fellowship (DE240100587). F.H.P. is supported by a Forrest Research Foundation Fellowship.

\section*{Data Availability}
Data sharing is not applicable to this article as no new data were created or analysed in this study.

\bibliographystyle{apsrev4-1}
\bibliography{refs}


\end{document}